\documentclass[journal]{IEEEtran}
\ifCLASSINFOpdf
\else
\fi
\usepackage{graphics} 
\usepackage{epsfig} 

\usepackage{times} 
\usepackage{amsmath,amsthm} 
\usepackage{amssymb}  
\usepackage{verbatim}
\usepackage{cases}
\usepackage{caption}
\usepackage{subcaption}
\usepackage{algorithm}
\usepackage[noend]{algpseudocode}
\usepackage{mathtools}
\usepackage{color}

\usepackage{bookmark}

\theoremstyle{definition}

\newtheorem{prob}{\textbf{Problem}}

\newtheorem{thm}{\textbf{Theorem}}

\newtheorem{corollary}{\textbf{Corollary}}

\newtheorem{defn}{\textbf{Definition}}

\newtheorem{exmp}{\textbf{Example}}

\newtheorem{prp}{\textbf{Proposition}}

\newtheorem{remk}{\emph{Remark}}

\newcommand{\arena}{$\mathcal{A}$}
\newcommand{\arenam}{\mathcal{A}}

\newcommand{\lang}{\mathcal{L}}
\newcommand{\control}{\mathcal{C}}

\begin{document}

\graphicspath{{Fig/}}

%
\title{Synthesis of Supervisors Robust Against Sensor Deception Attacks}
%
%
%

\author{R\^omulo Meira-G\'oes~\IEEEmembership{Student, IEEE},
		St\'ephane Lafortune~\IEEEmembership{Fellow, IEEE},
		Herv\'e Marchand
\thanks{The work of R.Meira-G\'oes and S. Lafortune was supported in part by US NSF grants CNS-1421122, CNS-1446298 and CNS-1738103.}
\thanks{R. Meira-G\'oes and S. Lafortune are with the Department of EECS, University of Michigan, MI 45109 USA (e-mail:\{romulo,stephane\}@umich.edu).}
\thanks{H. Marchand is with INRIA, Centre Rennes - Bretagne Atlantique, 35042 France (e-mail:herve.marchand@inria.fr).}
\thanks{\copyright 2020 IEEE. Personal use of this material is permitted.  Permission from IEEE must be obtained for all other uses, in any current or future media, including reprinting/republishing this material for advertising or promotional purposes, creating new collective works, for resale or redistribution to servers or lists, or reuse of any copyrighted component of this work in other works.}}

%
%

\markboth{}%
{ \MakeLowercase{\textit{}}: }
%



\maketitle

\begin{abstract}
We consider feedback control systems where sensor readings may be compromised by a malicious attacker intending on causing damage to the system.
We study this problem at the supervisory layer of the control system, using discrete event systems techniques.
We assume that the attacker can \emph{edit} the outputs from the sensors of the system before they reach the supervisory controller.
In this context, we formulate the problem of synthesizing a supervisor that is \emph{robust} against the class of edit attacks on the sensor readings and present a solution methodology for this problem.
This methodology blends techniques from games on automata with imperfect information with results from supervisory control theory of partially-observed discrete event systems.
Necessary and sufficient conditions are provided for the investigated problem.
\end{abstract}


%
\IEEEpeerreviewmaketitle

Protection of feedback control systems against cyber-attacks in critical infrastructures is an increasingly important problem.
In this paper, we consider sensor deception attacks at the supervisory layer of a feedback control system.
We assume that the underlying cyber-physical system has been abstracted as a discrete transition system (the \emph{plant} in this work), where sensor outputs belong to a finite set of (observable) events.
These events drive the supervisory controller, or simply \emph{supervisor}, that controls the high-level behavior of the system via actuator commands, which also belong to a finite set of (controllable) events.
In the context of this event-driven model, we incorporate a malicious \emph{attacker} that has compromised a subset of the observable events and is able to \emph{delete} actual sensor readings or to \emph{inject} fictitious ones in the communication channel to the supervisor.
The goal of the attacker is to leverage its knowledge of the plant and the supervisor models, and to use its event-editing capabilities to steer the plant state to a \emph{critical} state where damage to the plant occurs.
In this work, we investigate the problem of synthesizing a supervisor \emph{robust} against any attacker with these capabilities.

Several works have addressed in recent years problems of cyber-security in the above context.
In \cite{Thorsley:2006,Carvalho:2018,Lima:2019}, the authors developed diagnostic tools to detect when controlled systems are being attacked.
Their work is closely related to the work on fault diagnosis in discrete event systems, and it is applicable to both sensor and/or actuator attacks.
Our problem differs from the problem considered in these works since we aim to compute a supervisor that is robust against attacks without using a separate diagnostic tool. 
However, their method only works for attacks that are detectable/diagnosable (non-stealthy).
Moreover, once an attack is detected, their solution forces the supervisor to disable all controllable events.

%
There is also a vast literature in robust control in discrete event systems \cite{Lin:1993,Cury:1999,Xu:2009,Takai:2004,Rohloff:2012,Alves:2014,Lin:2014}.
However, robustness in the previous literature is related to communication delays \cite{Rohloff:2012,Alves:2014}, loss of information \cite{Lin:2014}, or model uncertainty \cite{Lin:1993,Cury:1999,Takai:2004,Xu:2009}.
Exceptions to that are \cite{Wakaiki:2017,Su:2018,Lin:2019cdc,Meira-Goes:2019c}, where the problem of synthesizing supervisors robust against attacks was investigated.
The results of \cite{Lin:2019cdc} are related to actuator deception attacks.

In \cite{Wakaiki:2017,Su:2018,Lin:2019cdc,Wang:2019a,Meira-Goes:2019c}, the problem of synthesizing supervisors robust against attacks was investigated.
Our work differs from \cite{Wakaiki:2017,Su:2018,Meira-Goes:2019c} as we provide a general game-theoretical framework that solves the problem of synthesizing supervisors robust against general classes of sensor deception attacks.
The solution methodology in \cite{Wakaiki:2017,Su:2018,Meira-Goes:2019c} follows the standard supervisory control solution methodology, where only results about \emph{one} robust supervisor against a \emph{specific class} of sensor deception attacks is provided. 
Conditions on the existence of robust supervisors against a possible set of sensor deception attacks with a normality condition on the plant are provided in \cite{Wakaiki:2017}.
A methodology to synthesize the supremal controllable and normal robust supervisor against \emph{bounded} sensor deception attacks is given in \cite{Su:2018}.
The results of \cite{Lin:2019cdc} are related to actuator and sensor replacement deception attacks while actuator and sensor deception attacks are considered in \cite{Wang:2019a}.
However, the supervisory control framework in \cite{Wang:2019a} differs from the standard framework since the authors assume that the supervisor can \emph{actively} change the state of the physical process.
Finally, \cite{Meira-Goes:2019c} provides a methodology to synthesize a maximal controllable and observable supervisor against \emph{unbounded} sensor deception attacks.

The game-theoretical framework adopted in this paper provides necessary and sufficient conditions for the problems of existence and synthesis of robust supervisors against general classes of sensor deception attacks.
This game-theoretical approach provides a structure that incorporates all robust supervisors against sensor deception attacks.
Different robust supervisors can be extracted from this structure, e.g., maximal controllable and observable, supremal controllable and normal, etc.
In fact, the robust supervisors from \cite{Wakaiki:2017,Meira-Goes:2019c} are embedded in this structure.
Moreover, there is a natural extension of our solution methodology such that robust supervisors from \cite{Su:2018} are embedded in this structure as well.

In summary, our work does not impose any normality condition as imposed in \cite{Wakaiki:2017,Su:2018} and studies synthesis and existence of robust supervisors against \emph{any} sensor deception attack.
Our approach considers both bounded and unbounded sensor deception attacks. 
Moreover, \emph{necessary and sufficient} conditions are provided for the existence and synthesis of robust supervisors, whereas in \cite{Wakaiki:2017} only existence conditions are provided and in \cite{Su:2018} only a sufficient condition is provided.

Of particular relevance to this paper is the work in \cite{Meira-Goes:2019a}, where the synthesis of stealthy sensor deception attacks assuming a fixed and known supervisor is considered; in this sense, \cite{Meira-Goes:2019a} pertains to \emph{attack} strategies.
Herein, we consider the ``dual'' problem of synthesizing a supervisor that is robust against sensor deception attacks; thus, this paper is focused on \emph{defense} strategies.

We wish to synthesize a supervisor that provably prevents the plant from reaching a critical state despite the fact that the information it receives from the compromised sensors may be inaccurate.
Our problem formulation is based on the following considerations.
The attack strategy is a parameter in our problem formulation, i.e., our problem formulation is parameterized by different classes of sensor deception attacks.
If there is no prior information about the attack strategy, then an ``all-out'' attack strategy is considered. 
Our solution methodology comprises two steps and leverages techniques from games on automata under imperfect information and from supervisory control of partially-observed discrete event systems.
We build a \emph{game arena} to capture the interaction of the attacker and the supervisor, under the constraints of the plant model.
The arena defines the solution space over which the problem of synthesizing supervisors with the desired robustness properties can be formulated.
In this solution space, called \emph{meta-system}, we use supervisory control techniques to enforce such robustness properties. 
We leverage the existing theory of supervisory control under partial observation \cite{Ramadge:1987,Lin:1988,Cieslak:1988,Cho:1989} to solve this \emph{meta-supervisory control problem}.
As formulated, the meta-supervisory control problem has a unique solution.
This solution embeds \emph{all} robust supervisors for the original plant, thereby providing a complete characterization of the problem addressed in this paper.

Our presentation is organized as follows.
Section \ref{sec:prem} introduces necessary background and the notation used throughout the paper. 
In Section \ref{sec:robust}, we formalize the problem of synthesis of supervisors robust against this attack model. 
We define the construction of the game arena and present the solution of the (meta-)synthesis problem in Section \ref{sec:meta}.
Section \ref{sec:supselect} discusses some benefits of our solution methodology.
Finally, we conclude the paper in Section \ref{sec:conclusion}.

\section{Preliminaries}\label{sec:prem}
We assume that the given cyber-physical system has been abstracted as a discrete transition system that we model as a finite-state automaton. 
A finite-state automaton $G$ is defined as a tuple $G = (X_G,\Sigma,\delta_G,x_{0,G})$, where $X_G$ is the finite set of states; $\Sigma$ is the finite set of events; $\delta_G: X_G\times \Sigma \rightarrow X_G$ is the partial transition function; $x_{0,G} \in X_G$ is the initial state.
The function $\delta_G$ is extended in the usual manner to domain $X_G\times\Sigma^*$. 
The language generated by $G$ is defined as $\mathcal{L}(G) = \{s \in \Sigma^*| \delta_G(x_{0,G}, s)!\}$, where ! means ``is defined".

In the context of supervisory control of DES \cite{Ramadge:1987}, system $G$ needs to be controlled in order to satisfy safety and liveness specifications.
In this work, we consider only safety specifications.
In order to control $G$, the event set $\Sigma$ is partitioned into the set of controllable events and the set of uncontrollable events, $\Sigma_{c}$ and $\Sigma_{uc}$.
The set of admissible admissible control decisions is defined as $\Gamma = \{\gamma\subseteq\Sigma|\Sigma_{uc} \subseteq \gamma\}$.
A supervisor, denoted by $S$, dynamically disables events such that the controlled behavior is provably ``safe''.
In other words, $S$ only disables controllable events to enforce the specification on $G$.

In addition, when the system is partially observed due to limited sensing capabilities of $G$,
the event set is also partitioned into $\Sigma = \Sigma_o\cup\Sigma_{uo}$, where $\Sigma_o$ is the set of observable events and $\Sigma_{uo}$ is the set of unobservable events. 
Based on this second partition, the \textit{projection} function $P_{\Sigma\Sigma_o}:\Sigma^*\rightarrow\Sigma_o^*$ is defined for $s\in \Sigma_o^*$ and $e\in\Sigma$ recursively as: $P_{\Sigma\Sigma_o}(\epsilon) = \epsilon$ and $P_{\Sigma\Sigma_o}(se) = P_{\Sigma\Sigma_o}(s)e$ if $e\in \Sigma_o$, $P_{\Sigma\Sigma_o}(s)$ otherwise. 
The inverse projection $P^{-1}_{\Sigma\Sigma_o}:\Sigma_o^*\rightarrow2^{\Sigma^*}$ is defined as $P^{-1}_{\Sigma\Sigma_o}(t) = \{s\in\Sigma^*|P_{\Sigma\Sigma_o}(s) = t\}$.

Supervisor $S$ makes its control decisions based on strings of observable events. 
Formally, a partial observation supervisor is a (partial) function $S:\Sigma_o^*\rightarrow\Gamma$. 
The resulting controlled behavior is a new DES denoted by $S /G$, resulting in the closed-loop language $\mathcal{L}(S/G)$, defined in the usual manner (see, e.g., \cite{Lafortune:2008}).
Normally, a supervisor $S$ is encoded by an automaton $R$ known as the supervisor realization, where every state encodes a control decision.
Throughout the paper, we use interchangeably supervisor $S$ and its realization $R$.

We also recall the notions of controllability, observability, and normality for a prefix-closed language $K\subseteq\mathcal{L}(G)$. We say the language $K$ is
\begin{itemize}
\item controllable w.r.t.\ to $\Sigma_c$, if $K\Sigma_{uc}\cap\mathcal{L}(G) \subseteq K$; 
\item observable w.r.t.\ to $\Sigma_o$ and $\Sigma_c$, if ($\forall s\in K, \forall e \in \Sigma_c: se \in K)[P_{\Sigma\Sigma_o}^{-1}(P_{\Sigma\Sigma_o}(s))e\cap\mathcal{L}(G)\subseteq K]$; 
\item normal w.r.t.\ to $\Sigma_o$ and $\Sigma_c$, if $K = P_{\Sigma\Sigma_o}^{-1}(P_{\Sigma\Sigma_o}(K))\cap \mathcal{L}(G)$.
\end{itemize}
\begin{exmp}\label{ex:supervisor}
We use the following example as illustrative example throughout the paper.
The plant $G$ is depicted in Fig.~\ref{fig:toy}(\subref{fig:grid1}) where $\Sigma_c = \Sigma_o = \{a,b\}$.
The supervisor shown in Fig.~\ref{fig:toy}(\subref{fig:robot_model}) guarantees that state $4$ is unreachable in the supervised system $R/G$.

\begin{figure}[thpb]
	\centering
	\begin{subfigure}[t]{.45\columnwidth}
		\centering
      \includegraphics[width=.55\columnwidth]{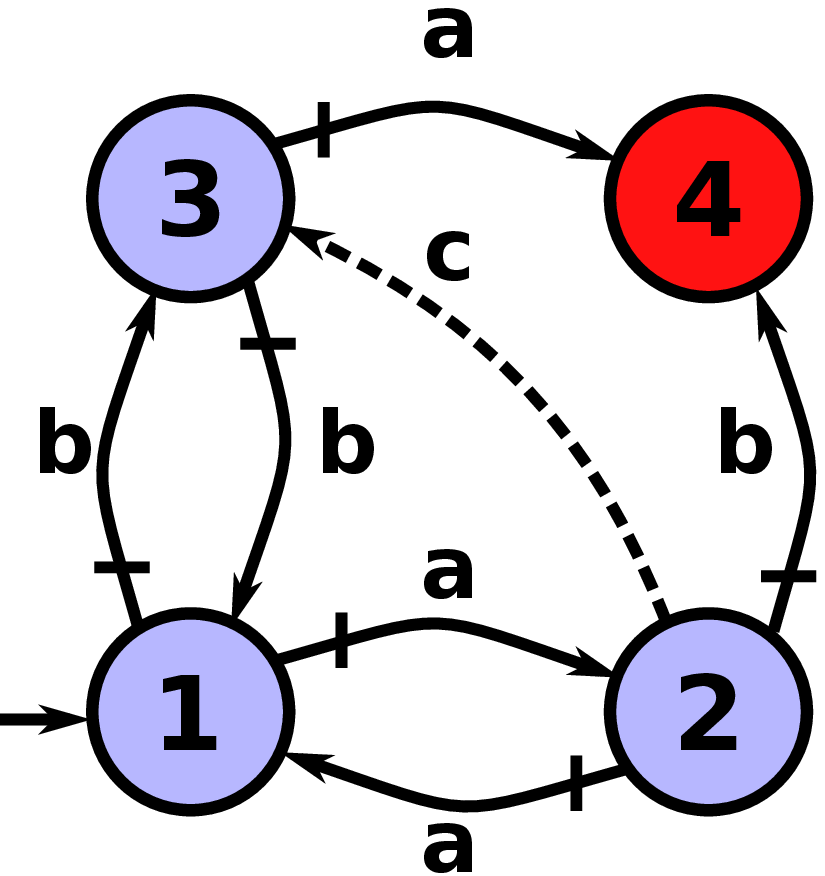}
      \caption{Plant $G$ with $\Sigma_c=\{a,b\}$ and $\Sigma_{o} = \{a,b\}$.}
      \label{fig:grid1}
	\end{subfigure}%
	\quad
	\begin{subfigure}[t]{.45\columnwidth}
		\centering
      \includegraphics[width=.52\columnwidth]{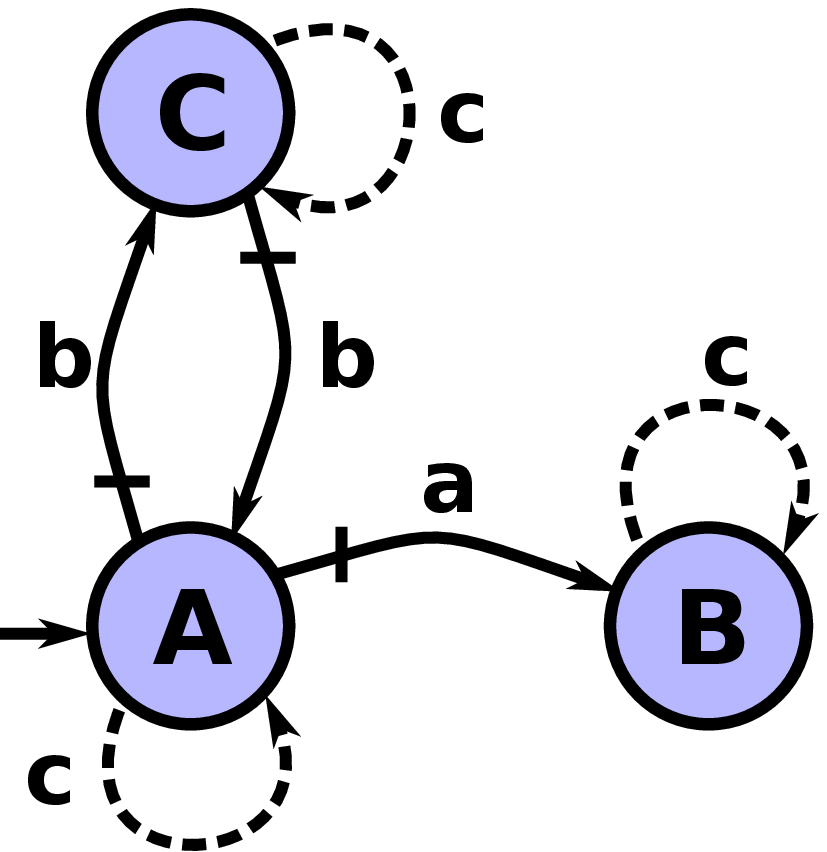}
      \caption{Supervisor $R$.}
      \label{fig:robot_model}
	\end{subfigure}
	\caption{Running example. Observable events have solid arrows and unobservable events have dashed arrows. Controllable events have marks across their arrows.}
	\label{fig:toy}
\end{figure}
\end{exmp}

For convenience, we define useful operators and notation that we  use throughout this paper.
First, $\Gamma_G(Q)$ is defined as the set of active events at the set of states  $Q \subseteq X_G$ of the automaton $G$, given by:
\begin{align}
\Gamma_G(Q) := \{e\in\Sigma|(\exists x \in Q)[\delta_G(x,e)!]\}
\end{align}
By an abuse of notation, we use $\Gamma_G(x) = \Gamma_G(\{x\})$ for $x\in X_G$.
 
The unobservable reach of the subset of states $Q \subseteq X_G$ under the subset of events $\gamma\subseteq\Gamma$ is given by:
\begin{align}\label{eq:UR}
UR_\gamma(Q) &:= \{x\in X_G \mid (\exists t \in (\Sigma_{uo}\cap\gamma)^*)[x \in \delta_G(Q,t)]\}
\end{align}
where $\delta_G(Q,t) = \cup_{x\in Q}\{\delta_G(x,t)\}$ and we consider $\delta_G(x,t) = \emptyset$ if $\delta_G(x,t)$ is not defined.
The observable reach of the subset of states $Q\subseteq X_G$ given the execution of the observable event $e\in\Sigma_o$ is defined as:
\begin{equation}\label{eq:nx}
NX_e(Q) := \delta_G(Q,e)
\end{equation}

We define by $trim(G,Q)$ the operation that returns the \textit{accessible} subautomaton of $G$ after deleting states $Q\subseteq X_G$.
For any string $s\in \Sigma^*$, $|s|$ is the length of $s$. 
We denote by $e^i_s$ the $i^{th}$ event of $s$ such that $s = e^1_se^2_s\ldots e^{|s|}_s$.
Lastly, $s^i$ denotes the $i^{th}$ prefix of $s$, i.e., $s^i = e^1_s\ldots e^i_s$ and $s^0 = \epsilon$.  

\section{Robust Supervisory Control against Deception Attacks}\label{sec:robust}

\subsection{Notation}
We first define useful notation for this section.
Since we consider that the observability properties of the events are static, and not dynamic, we assume that the attacker only affects observable events; clearly, an insertion of an unobservable event would lead to immediate detection of the attacker by the supervisor (whose transition function is only defined for observable events).
For this reason, we define the set $\Sigma_a\subseteq\Sigma_o$ to be the compromised event set.
These are the events that the attacker has the ability to alter, where ``alter" means it can insert or delete events.

We define the set of inserted events $\Sigma_a^i = \{e_i\mid e \in \Sigma_a\}$ and the set of deleted events $\Sigma_a^d = \{e_d\mid e \in \Sigma_a\}$. 
These sets represent the actions of an attacker, and we use subscripts to distinguish them from events generated by $G$ such that $\Sigma^i_a\cap\Sigma = \Sigma^d_a\cap\Sigma  = \Sigma^i_a\cap\Sigma^d_a  = \emptyset$.
We call the events in $\Sigma$ as legitimate events, events that are not insertion nor deletion.
For convenience, we define $\Sigma_a^e = \Sigma_a^i\cup\Sigma_a^d$, $\Sigma_{o,e} = \Sigma_o\cup\Sigma_a^e$ and $\Sigma_m = \Sigma \cup \Sigma_a^e$. 

We define three projection operators with $\Sigma_m$ as domain and $\Sigma$ as co-domain: 
(1) $\mathcal{M}$ is defined as $\mathcal{M}(e_i) = \mathcal{M}(e_d) = \mathcal{M}(e) = e$ for $e\in\Sigma$; (2) $P^G(e) = \mathcal{M}(e)$ for $e\in\Sigma\cup\Sigma^d_a$ and $P^G(e)=\epsilon$ for $e\in\Sigma_a^i$;
(3) $ P^S(e) = \mathcal{M}(e)$ for $e\in\Sigma\cup\Sigma_a^i$ and $P^S(e)=\epsilon$ for $e\in\Sigma^d_a$.
The mask $\mathcal{M}$ removes subscripts, when present, from events in $\Sigma_m$, $P^G$ projects an event in $\Sigma_m$ to its actual event execution in $G$, and $P^S$ projects an event in $\Sigma_m$ to its event observation by $S$.

%

\subsection{Modeling sensor deception attacks}

We assume that the attacker hijacks the communication channel between the plant and the supervisor and it can modify the readings of events in $\Sigma_a$, as depicted in Fig.~\ref{fig:model_cdc}.
Intuitively, the attacker is modeled similarly as a supervisor.
The attacker takes its actions based on observing a \emph{new event} $e\in \Sigma_o$ from $G$ and its memory of the \emph{past modified string}.
Note that, we assume that the attacker observes the same observable events as the supervisors.
Formally, we model an attacker as a \emph{nondeterministic} string edit function.
\begin{defn}\label{def:attacker}
Given a system $G$ and a subset $\Sigma_a\subseteq\Sigma_o$, an attacker is defined as a partial function $f_A:\Sigma_{o,e}^* \times (\Sigma_{o}\cup \{\epsilon\}) \rightarrow 2^{\Sigma_{o,e}^*}\setminus\emptyset$ s.t. $f_A$ satisfies the following constraints $\forall s\in \Sigma_{o,e}^*$ and $e\in \Sigma_o$:
\begin{enumerate}
\item $f_A(\epsilon,\epsilon) \subseteq{\Sigma_a^i}^*$; $f_A(s,\epsilon)=\{\epsilon\}$ when $s\neq \epsilon$;
\item If $e\in\Sigma_o \setminus \Sigma_a $: $f_A(s,e) \subseteq\ \{e\}{\Sigma_a^i}^*$;
\item If $e\in\Sigma_a$: $f_A(s,e) \subseteq \ \{e,\ e_d\}{\Sigma_a^i}^*$.
\end{enumerate}
\end{defn}

The function $f_A$ captures a general model of deception attack.
Namely, $f_A$ defines a substitution rule where the observation $e$ is replaced \emph{by a string} in the set $f_A(s,e)$.
Condition (1) allows event insertions when the plant is in the initial state and constrains the substitution rule based on observation of events from $G$\footnote{Observe that clause (1) of Def.~\ref{def:attacker} corrects a mistake in the corresponding clause (1) of Def.~2 in \cite{Meira-Goes:2019a}, where $\emptyset$ was inadvertently used instead of $\{ \epsilon \}$ for initializing $f_A (s,\epsilon)$.}.
Condition (2) constrains the attacker from erasing $e$ when $e$ is outside of $\Sigma_a$. 
However, the attacker may insert an arbitrary string $t\in{\Sigma^i_a}^*$ \emph{after} the occurrence of $e$. 
Lastly, condition (3) allows events $e\in\Sigma_a$ to be edited to any string $t\in\{e,\ e_d\}{\Sigma_a^i}^*$.

For simplicity, we assume that the function $f_A$ has been encoded into a finite-state automaton $A = (X_A,\Sigma_{o,e},\delta_A,x_{0,A})$ where $\delta_A$ is complete with respect to $\Sigma_o\setminus\Sigma_a$ and for any $(e\in\Sigma_a,\ q\in X_A)$ then $(\delta_A(q,e)!\vee\delta_A(q,e_d)!)$.
This assumption will be used later when we explain the composition in the definition of the closed-loop behavior under attack.
Let $A$ encode an $f_A$, then the function $f_A$ is extracted from $A$ as follows: $\forall s \in \lang(A)$ and $e\in\Sigma_o$, $f_A(s,e) = \{ t\in\{e,e_d\}{\Sigma_a^i}^*\mid \delta_A(x_{0,A},st)!\}$, $f_A(\epsilon,\epsilon) = \{ t\in {\Sigma_a^i}^*\mid \delta_A(x_{0,A},t)!\}$, and $f_A(s,e)$ is undefined for all $s\in \Sigma_{o,e}^*\setminus\lang(A)$ and $e\in \Sigma_o$. 
In Appendix~\ref{app:attack}, we show how to relax the above assumption on automaton $A$ to encode attack functions.

This formulation provides a simple way to handle attack functions and it characterizes the behavior of the attacker.
It also provides a way to define specific attackers that are more constrained than the constraints of Definition~\ref{def:attacker}, i.e., when some prior knowledge about the attacker is available.
In other words, the automaton $A$ can encode different attack strategies, e.g., replacement attack, bounded attack, etc.

One important attack strategy for this problem is the “all-out” attack strategy introduced in \cite{Carvalho:2018,Lima:2017}.
In this model, the attacker could attack whenever it is possible.
Hereafter, if there is no prior information about the attack strategy, then we assume that the attacker follows the all-out attack strategy.
The following example provides two attack strategies for Example~\ref{ex:supervisor}, one of these strategies is the all-out strategy.

\begin{exmp}
Attack functions $f_{A^1}$ and $f_{A^2}$ for the system defined in Example~\ref{ex:supervisor} and $\Sigma_a = \{b\}$ were encoded in automata $A_1$ and $A_2$ depicted in Fig.~\ref{fig:attacks}.
Automaton $A^1$ encodes the all-out strategy for this example.
Although the all-out strategy is a nondeterministic strategy since the attacker can try all possible combinations of attacks, its automaton representation is a deterministic automaton.
Only one state is necessary to encode the all-out strategy. 
Automaton $A^2$ encodes a one sensor reading deletion attack strategy.

\begin{figure}[thpb]
	\centering
	\begin{subfigure}[t]{.45\columnwidth}
		\centering
      \includegraphics[width=.35\columnwidth]{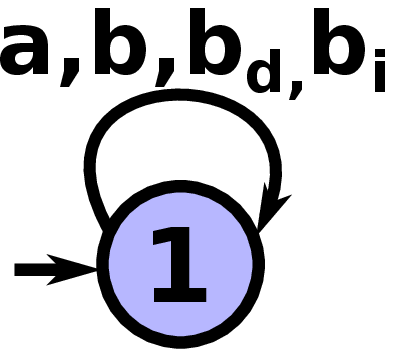}
      \caption{$A^1$ - all-out strategy.}
      \label{fig:attack01}
	\end{subfigure}%
	\quad
	\begin{subfigure}[t]{.45\columnwidth}
		\centering
      \includegraphics[width=0.6\columnwidth]{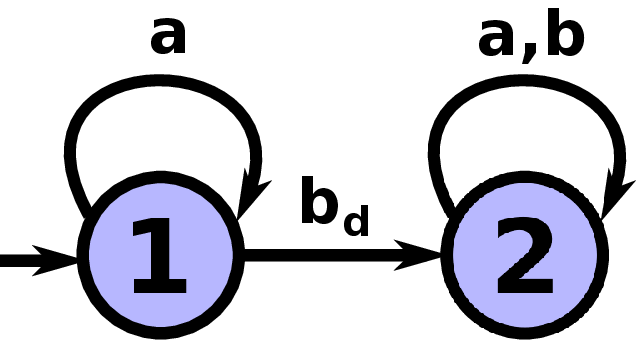}
      \caption{$A^2$ - one deletion strategy}
      \label{fig:attack02}
	\end{subfigure}
	\caption{Representation of two attack functions}
	\label{fig:attacks}
\end{figure}
\end{exmp}

\subsection{Controlled system under sensor deception attack}

\begin{figure}[thpb]
      \centering
      \includegraphics[width=.55\columnwidth]{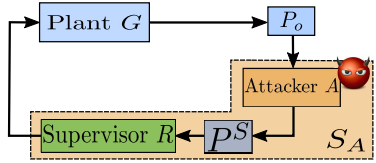}
      \caption{Sensor deception attack framework}
      \label{fig:model_cdc}
\end{figure}

The attacker in the controlled system induces a new controlled language.
Referring to Fig.~\ref{fig:model_cdc}, $R$, $A$ and $P^S$ together effectively generate a new supervisor $S_A$ for the system $G$.

To characterize the interaction of attacker $A$ with the system $G$ and supervisor realization $R$, we must modify the behavior of $G$ and $R$ such that it takes into account possible modifications of $A$.
We use the method in \cite{Meira-Goes:2019c}, where $G$ and $R$ are augmented with attack actions providing an attacked system $G_a$ and an attacked supervisor $R_a$.

\begin{defn}\label{def:G_a}
Given $G$ and $\Sigma_a$, we define the attacked plant $G_a$ as: $G_a = (X_{G_a} = X_{G},\Sigma_m = \Sigma\cup\Sigma_a^e,\delta_{G_a},x_{0,G_a} = x_{0,G})$ where $\delta_{G_a}(x,e) = \delta_G(x,P^G(e))$ and $\delta_G(x,\epsilon) = x$.
\end{defn}

Similarly to the construction of $G_a$, we can modify the behavior of $R$ to reflect the modifications made by an attacker on the communication channel.

\begin{defn}\label{def:R_a}
Given $R$ and $\Sigma_a$, we define the attacked supervisor as: $R_a=(X_{R_a} = X_{R},\Sigma_m,\delta_{R_a},x_{0,R_a} = x_{0,R})$ where $$\delta_{R_a}(x,e) :=\left\{ 
\begin{array}{ll} 
\delta_R(x,P^S(e))&\text{if } \mathcal{M}(e)\in \Gamma_R(x)\\
x & \text{if } e\in \Sigma_a^i \text{ and } \mathcal{M}(e)\not\in \Gamma_R(x)\\
\text{undefined} & \text{otherwise}
\end{array}\right.$$
\end{defn}

We assume that the supervisor ``ignores" insertions of controllable events that are not enabled by the current control action at state $x$.
This assumption is specified by the second condition in the definition of $\delta_{R_a}$.
Namely, the insertion made by the attacker is ineffective at this state.
In some sense, this means that the supervisor ``knows'' that this controllable event has to be an insertion performed by the attacker, since it is not an enabled event.

Based on $G_a$, $R_a$ and $A$, we define the closed-loop language of the attacked system to be $\lang(S_A/G) = P^G(\lang(G_a||R_a||A))$, where $||$ is the standard parallel composition operator \cite{Lafortune:2008}.
Recall that the transition function of $A$ is complete with respect to $\Sigma_o\setminus\Sigma_a$ and for any $(e\in\Sigma_a,\ q\in X_A)$ then $\delta_A(q,e)!\vee\delta_A(q,e_d)!$.
Therefore, the attacker is incapable of disabling events of $G$.  

\begin{exmp}
We return to our running example. 
Figure~\ref{fig:attacked} depicts the attacked system $G_a$, the attacked supervisor $R_a$, and the supervised attacked system $G_a||R_a||A^1$, where $A^1$ is the all-out attack strategy shown in Fig.~\ref{fig:attacks}(\subref{fig:attack01}).
Note that state $4$ is reachable in the supervised attacked system. 
\begin{figure}[thpb]
	\centering
	\begin{subfigure}[t]{.45\columnwidth}
		\centering
      \includegraphics[width=.65\columnwidth]{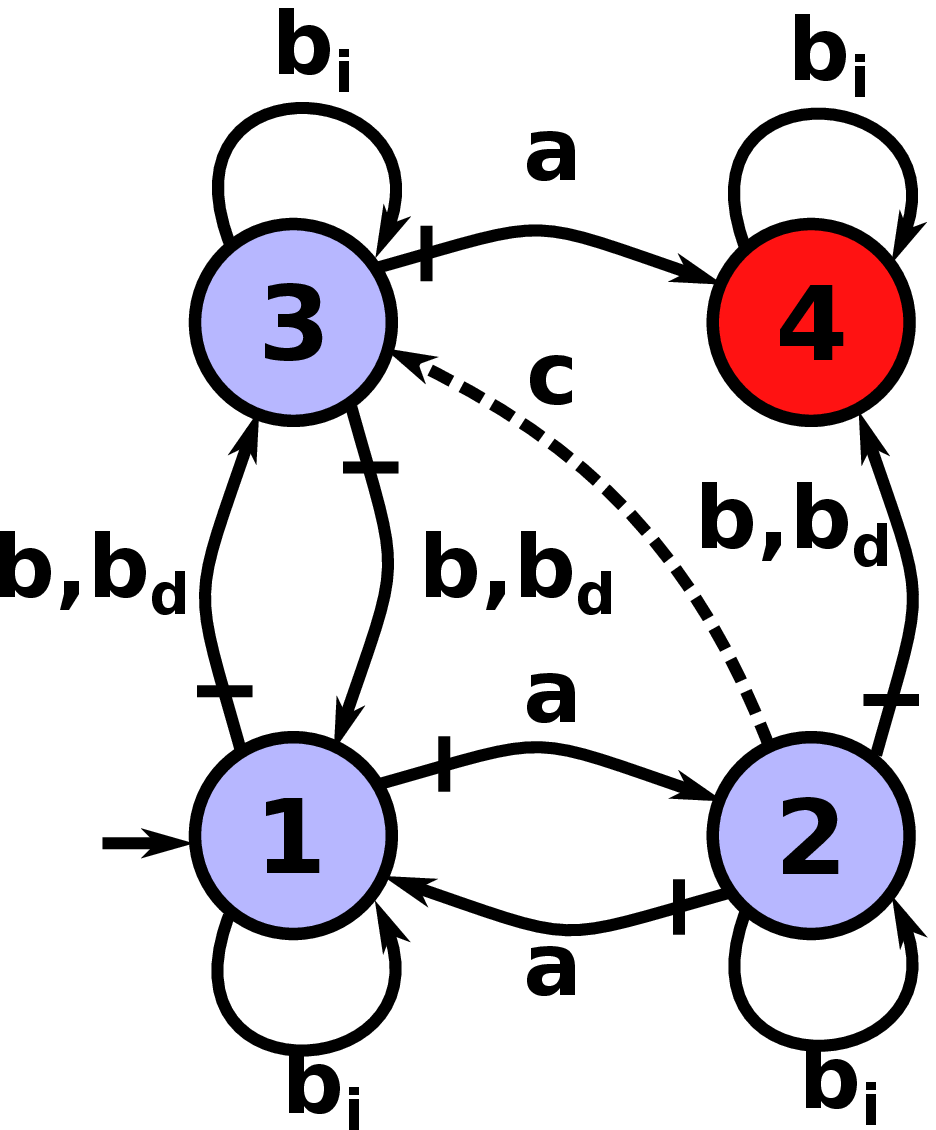}
      \caption{$G_a$}
      \label{fig:Ga}
	\end{subfigure}%
	\quad
	\begin{subfigure}[t]{.45\columnwidth}
		\centering
      \includegraphics[width=.75\columnwidth]{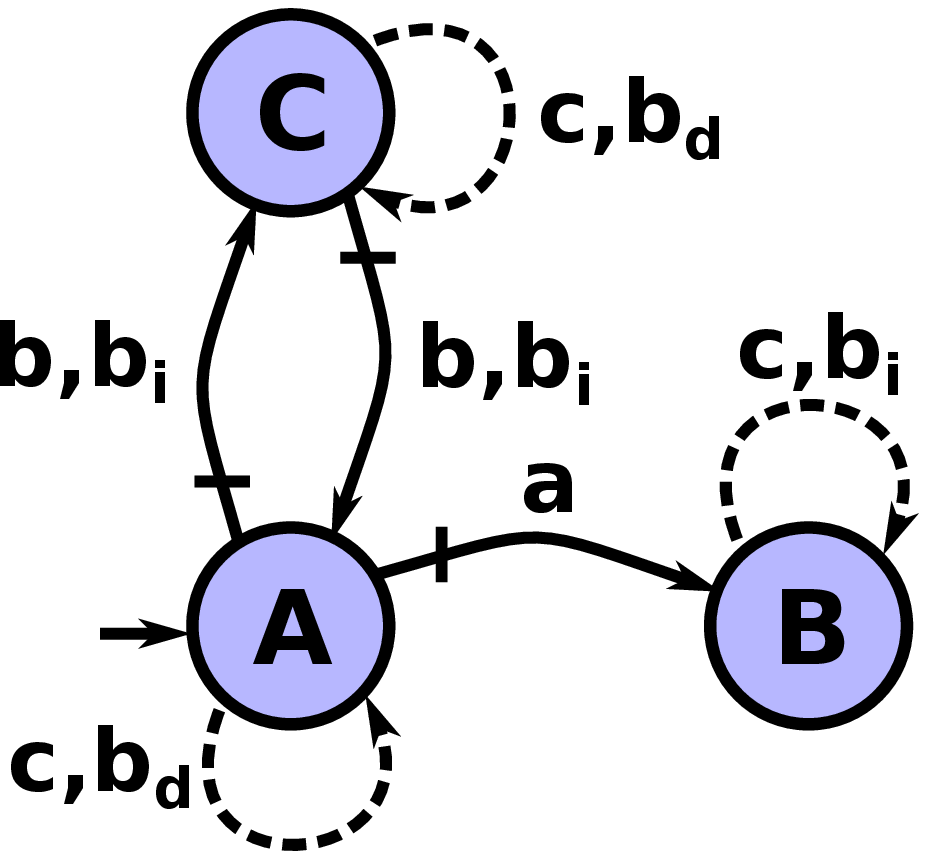}
      \caption{$R_a$}
      \label{fig:R_a}
	\end{subfigure}
	\quad
	\begin{subfigure}[t]{.6\columnwidth}
		\centering
      \includegraphics[width=.9\columnwidth]{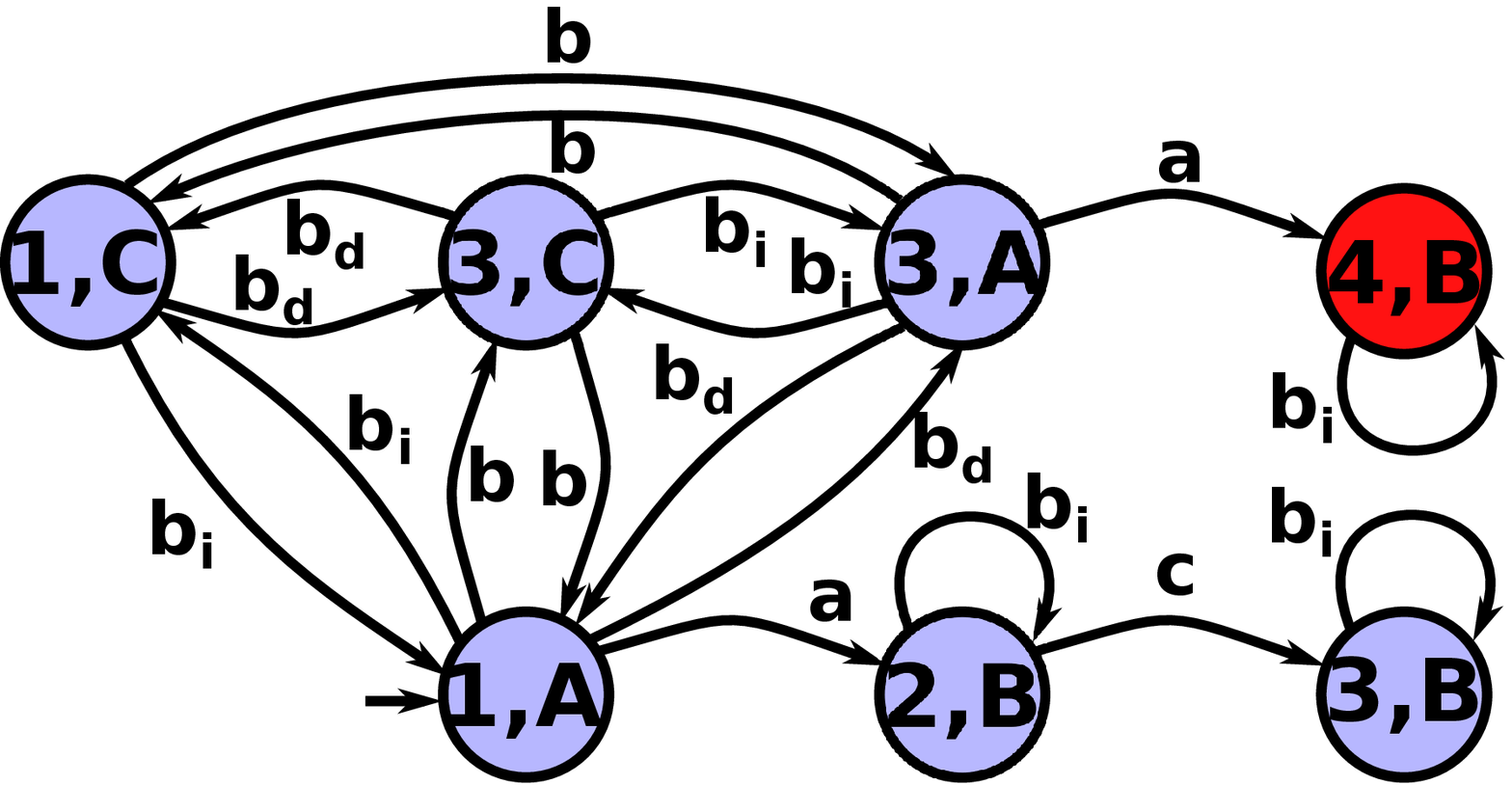}
      \caption{$G_a||R_a$}
      \label{fig:GaRa}
	\end{subfigure}
	\caption{Supervisory control under sensor deception attack}
	\label{fig:attacked}
\end{figure}
\end{exmp}

\begin{remk}
Even though the attack function $f_A$ is nondeterministic, the language generated by the attacked system is uniquely defined, i.e., $\lang(S_A/G) = P^G(\lang(G_a||R_a||A))$.
In \cite{Wakaiki:2017}, the nondeterministic attack function defined therein generates maximal and minimal attacked languages.
Similar to the problem encountered in \cite{Shu:2015}, the maximal language possibly contains strings that the supervised plant cannot generate while the minimal does not define all possible strings that this controlled plant generates.
This issue does not arise in our context.
Our language definition also differs from the one in \cite{Su:2018}.
Even though an attacker could have a string of insertions to send to the supervisor, it does so by sending one event at the time.
On the other hand in \cite{Su:2018}, the attacker sends the entire string modification to the supervisor.
\end{remk}

\subsection{Robustness against deception attacks}

We investigate the problem of synthesizing a supervisor $R$ robust against the attack strategy $A$.
We assume that the plant $G$ contains a set of \textit{critical} states defined as $X_{crit}\subset X_G$; these states are unsafe in the sense that they are states where physical damage to the plant might occur.
Although damage is defined in relation to the set $X_{crit}$, it could be generalized in relation to any regular language by state space refinement.

\begin{defn}\label{def:robustsup}
Supervisor $R$ is \emph{robust} (against sensor deception attacks) with respect to $G$, $X_{crit}$ and $A$, if for any $s \in \mathcal{L}(S_A/G)$ then $\delta_G(x_{0,G},s) \not\in X_{crit}$.
\end{defn}

The definition of robustness is dependent on the attack strategy $A$.
Recall that the all-out strategy encompasses all other attack strategies \cite{Carvalho:2018}.
Therefore, a supervisor that is robust against the all-out strategy is robust against any other $A$ \cite{Meira-Goes:2019c}. 

\begin{prob}[Synthesis of Robust Supervisor] \label{prob:synthesis}
Given $G$, $X_{crit}$ and an attack strategy $A$, synthesize a robust supervisor $R$, if one exists, with respect to $G$, $X_{crit}$ and $A$.
\end{prob}

We are asking that the robust supervisor should prevent the plant from reaching a critical state regardless of the fact that it might receive inaccurate information. 
In other words, the supervisor will react to every event that it receives, but since it was designed to be robust to $A$, the insertions and deletions that $A$ performs will never cause $G$ to reach $X_{crit}$.
This will be guaranteed by the solution procedure presented in the next section.

\section{Meta-Supervisor problem} \label{sec:meta}
\begin{figure}[thpb]
      \centering
      \includegraphics[width=.70\columnwidth]{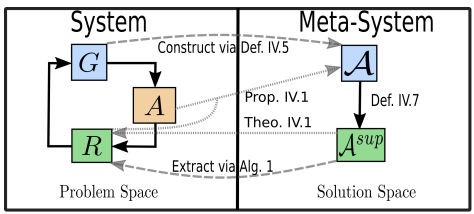}
      \caption{Relation of the system and the meta-system}
      \label{fig:relationsys-meta}
\end{figure}

In this section, we present our approach to solve Problem~\ref{prob:synthesis}.
We briefly explain the idea of our approach.
Figure~\ref{fig:relationsys-meta} shows the connection of the problem formulation space (left box) and the solution space (right box).
The connection between these two spaces is given by the arrows that cross the two boxes.
These arrows are labeled by results provided in this section.

In the left box of Fig. \ref{fig:relationsys-meta}, we have the problem formulation space where the supervisor $R$ is unknown.
Based on $G$, $\Sigma_o$, $\Sigma_c$ and $A$, we construct a \emph{meta-system}, called $\arenam$, in a space where all supervisors are defined. 
This construction is given in Definition~\ref{def:arena}.
The meta-system is part of the proposed solution space and it is represented in the right box of Fig.~\ref{fig:relationsys-meta}.

Although all supervisors are defined in $\arenam$, which is shown by Proposition~\ref{prop:runs_in_A}, we are only interested in robust supervisors.
In order to obtain robust supervisors, we use techniques of partially observed supervisory control theory \cite{Cieslak:1988,Cho:1989} in the meta-system.
The structure $\arenam^{\sup}$ is obtained via Definition~\ref{def:meta-control} and it contains all robust supervisors against sensor deception attacks on $\Sigma_a$.

Finally, to return to our problem formulation space, we extract one supervisor, if one exists, from $\arenam^{\sup}$.
Such extraction is given by Algorithm~\ref{algo:constructionRr}.

\subsection{Definition} \label{subsection:def}

Inspired by the techniques of two-player reachability games, we construct an arena as it is constructed in these games.
In the arena, player 1 represents the supervisor while player 2 represents the adversarial environment. 
The arena exhaustively captures the game between the supervisor and the environment, where the supervisor selects control decisions ($\Gamma$) and the environment executes events ($\Sigma_{o,e}$).
In the arena, player 1's transitions record a control decision made by the supervisor.
On the other hand, player 2's transitions represent actions of the plant $G$ or actions of the attacker $A$.
Formally, the arena is defined as follows.

\begin{defn}\label{def:arena}
Given plant $G$ and attack function $A$, we define the arena $\mathcal{A}$ as 4-tuple:
\begin{equation}
\mathcal{A} = (Q_1\cup Q_2,A_1\cup A_2,h_1\cup h_2,q_0)
\end{equation}
where,
\begin{itemize}
\item $Q_1 \subseteq 2^{X_G}\times{X_A}$ is the set of states where the supervisor issues a control decision. 
Its states have the form of $(S_1,S_2)$, where $S_1$ is the estimate of the states (as it is executed by the plant) of $G$ and $S_2$ is the attacker's state. 
For convenience we define the projection operators $I_i((S_1,S_2)) = S_i$ for $i \in \{1,2\}$;
\item $Q_2 \subseteq 2^{X_G}\times{X_A}\times\Gamma\times (\{\epsilon\}\cup\Sigma_a)$ is the set of states where the adversarial environment issues a decision. 
Its states have the form $(S_1,S_2,\gamma,\sigma)$, where $S_1$ and $S_2$ are defined as in $Q_1$ states, $\gamma$ is the last control decision made by the supervisor, and $\sigma$ is related to inserted events.
The event $\sigma$ is  equal to $e\in\Sigma_a$ if the last transition was $e_i \in \Sigma_a^i$, otherwise it is equal to $\epsilon$. 
We use the same projection operators $I_i$ for states in $Q_2$ for $i \in \{1,2\}$;
\item $A_1 = \Gamma$ and $A_2 = \Sigma_{o,e}$ are respectively the actions/decisions of player 1 and player 2;
\item $h_1:Q_1\times A_1\rightarrow Q_2$ is built as follows: for any $q_1=(S_1,S_2)\in Q_1$ and $\gamma\in A_1$
\begin{gather}\label{eq:h1}
\begin{split}
&h_1(q_1,\gamma) := \big(UR_{\gamma}(S_1),S_2,\gamma,\epsilon\big)
\end{split}
\end{gather}

\item $h_2:Q_2\times A_2\rightarrow Q_1\cup Q_2$ is built as follows for any $q_2=(S_1,S_2,\gamma,\sigma)\in Q_2$: 
\end{itemize}
Let $e \in \Sigma_o$:
\begin{align}\label{eq:h2_obs}
h_2(q_2,e) = \left\{\hspace*{-0.2cm}
				\begin{array}{ll}
				\big(NX_{e}(S_1),\delta_A(S_2,e)\big) &\text{if } (e \in \Gamma_G(S_1)\cap\gamma)\wedge\\& \hspace*{-0.5cm}\hfill(e\in\Gamma_A(S_2)) \wedge(\sigma = \epsilon) \\
				\big(S_1,S_2\big)& \text{if } (\sigma = e)\\
				\text{undefined}&  \text{otherwise} 	
\end{array}				 
\right.
\end{align}
Let $e\in \Sigma_a$:
\begin{align}\label{eq:h2_ins}
h_2(q_2,e_i) = \left\{\hspace*{-0.2cm}
				\begin{array}{ll}
				\big(S_1,\delta_A(S_2,e_i),\gamma,e\big)&\text{if } (e_i \in\Gamma_A(S_2))\wedge\\&\hfill(\sigma = \epsilon)\\
				\text{undefined}&  \text{otherwise} 	
\end{array}				 
\right.
\end{align}
\begin{align}\label{eq:h2_del}
h_2(q_2,e_d) = \left\{\hspace*{-0.2cm}
				\begin{array}{ll}
				\big(UR_{\gamma}(NX_{e}(S_1)),\delta_A(S_2,e_d),\gamma,\epsilon\big) &\\
				&\hfill\hspace*{-3cm}\text{if } (e\in \Gamma_G(S_1)\cap\gamma)\wedge\\
				&\hspace*{-3cm}\hfill(e_d \in\Gamma_A(S_2))\wedge (\sigma = \epsilon)\\
				\text{undefined}&\hspace*{-2cm}\hfill\text{otherwise} 	
\end{array}				 
\right.
\end{align}
\begin{itemize}
\item $q_0 \in Q_1$ is the initial S-state: $q_0 := (\{x_{0,G}\},x_{0,A})$.
\end{itemize}
\end{defn}

We explain the definition of the transition functions $h_1$ and $h_2$ in detail.
The definition of $h_1$ is simple and it defines a transition from player 1 to player 2, which records a control decision made by the supervisor, and it updates $G$'s state estimate according to this decision. 
On the other hand, $h_2$ is more complex since player 2 has two types of transitions.

\begin{figure}[thpb]
	\centering
	\begin{subfigure}[t]{.45\columnwidth}
		\centering
		\includegraphics[width=.7\columnwidth]{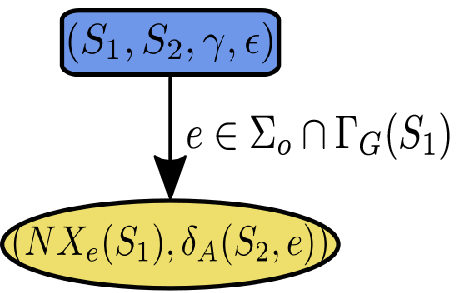}
		\caption{First transition of Eq.~(\ref{eq:h2_obs})}
		\label{fig:trans1}
	\end{subfigure}%
	\quad
	\begin{subfigure}[t]{.45\columnwidth}
		\centering
		\includegraphics[width=.48\columnwidth]{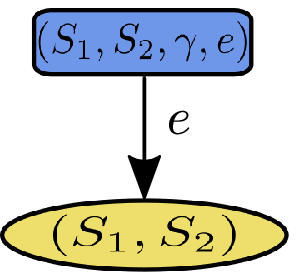}
		\caption{Second transition of Eq.~(\ref{eq:h2_obs})}
		\label{fig:trans2}
	\end{subfigure}
	\caption{Transition function $h_2$ from player 2 to player 1 }
	\label{fig:trans21}
\end{figure}

The first type is transitions from player 2 to player 1 which characterizes the visible decision made by the environment and is related to events in $\Sigma_o$.
These transitions are defined in Eq. (\ref{eq:h2_obs}), and they are illustrated in Fig.~\ref{fig:trans21}. 
In Fig.~\ref{fig:trans21}(\subref{fig:trans1}), an event $e\in\Sigma_o$ that is feasible in $G$ from some state in $S_1$ is selected; thus, both the state estimate and the attacker's state are updated. 
In Fig~\ref{fig:trans21}(\subref{fig:trans2}), $q_2 = (S_1,S_2,\gamma,e)\in Q_2$ is reached after an insertion since  $e \neq \epsilon$; thus, $G$'s state estimate and the attacker's state remain unchanged.

\begin{figure}[thpb]
	\centering
	\begin{subfigure}[t]{.43\columnwidth}
		\centering
		\includegraphics[width=.58\columnwidth]{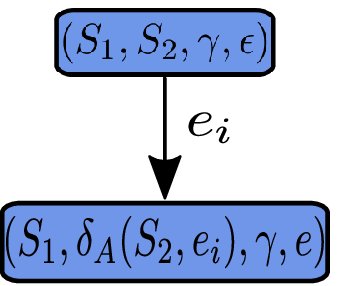}
		\caption{Transition of Eq.~(\ref{eq:h2_ins})}
		\label{fig:trans3}
	\end{subfigure}%
	\quad
	\begin{subfigure}[t]{.45\columnwidth}
		\centering
		\includegraphics[width=.95\columnwidth]{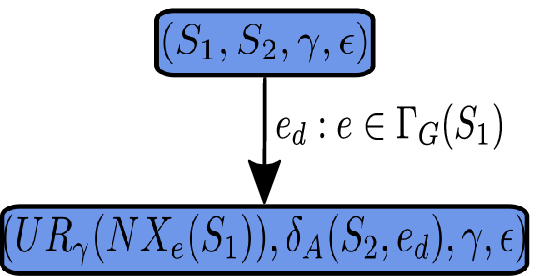}
		\caption{Transition of Eq.~(\ref{eq:h2_del})}
		\label{fig:trans4}
	\end{subfigure}
	\caption{Transition function $h_2$ from player 2 to player 2 }
	\label{fig:trans22}
\end{figure}

Transitions from player 2 to itself characterize invisible, from the supervisor's perspective, decisions. They are only defined for events in $\Sigma_a^e$ . 
These transitions are defined by Eqs.~(\ref{eq:h2_ins}-\ref{eq:h2_del}).
An attacker can insert any event in $e\in\Sigma_a$, as long as $e_i$ is allowed in the current attacker's state.
The inserted events $e\in\Sigma_a^i$ are not going to be seen by the supervisor, as only the attacker knows it decided to insert the event. 
Insertions will be seen by the supervisor as genuine events.
But $e_i$ represents here (in the context of the game arena) the intention of the attacker to insert.
Equation~(\ref{eq:h2_ins}) (depicted in Fig.~\ref{fig:trans22}(\subref{fig:trans3})) is the unobservable part, where an insertion decision was selected and the attacker's state and the fourth component of $q_2\in Q_2$ are updated. 
The observable part is shown by Fig.~\ref{fig:trans21}(\subref{fig:trans2}). 
In the case of a deleted event, from the supervisor's perspective, it is seen as an $\epsilon$ event as well. 
That is, the supervisor cannot change its control decision when the attacker deletes an event, as shown in Fig.~\ref{fig:trans22}(\subref{fig:trans4}).


\textit{Remark 1}: The elements of $Q_1$ and $Q_2$ are defined such that they incorporate the ``sufficient information'' (in the sense of \emph{information state} in system theory) that each player needs to make its respective decision. 
Equations~(\ref{eq:h1}-\ref{eq:h2_del}) guarantee by construction that the updates of the information states are consistent with the plant dynamics and the actions of the attacker.
Overall, the arena constructed thereby captures the possible attacks and all possible supervisors in a finite structure.
We prove both results later on.

\begin{exmp}\label{ex:partialarena}
We return to our illustrative example to show results on the construction of the arena. 
We construct \arena\ for the system $G$, $X_{crit} = \{4\}$ and $A^1$ depicted in Fig.~\ref{fig:attacks}(\subref{fig:attack01}).
Since we construct $\arenam$ for the all-out attack strategy, we can omit the attacker state. 
The arena has a total of $26$ states.
Figure~\ref{fig:arena} illustrates arena $\arenam$ constructed with respect to $A^1$ and $G$.
We can observe the encoding of insertion and deletion in this arena.
For example, at state $(\{1\},\{a,b,c\},\epsilon)$ the transition $b_i$ goes to state $(\{1\},\{a,b,c\},b)$ and then transition $b$ takes state $(\{1\},\{a,b,c\},b)$ to state $(\{1\})$.

\begin{figure*}[thpb]
      \centering
      \includegraphics[width=0.7\textwidth]{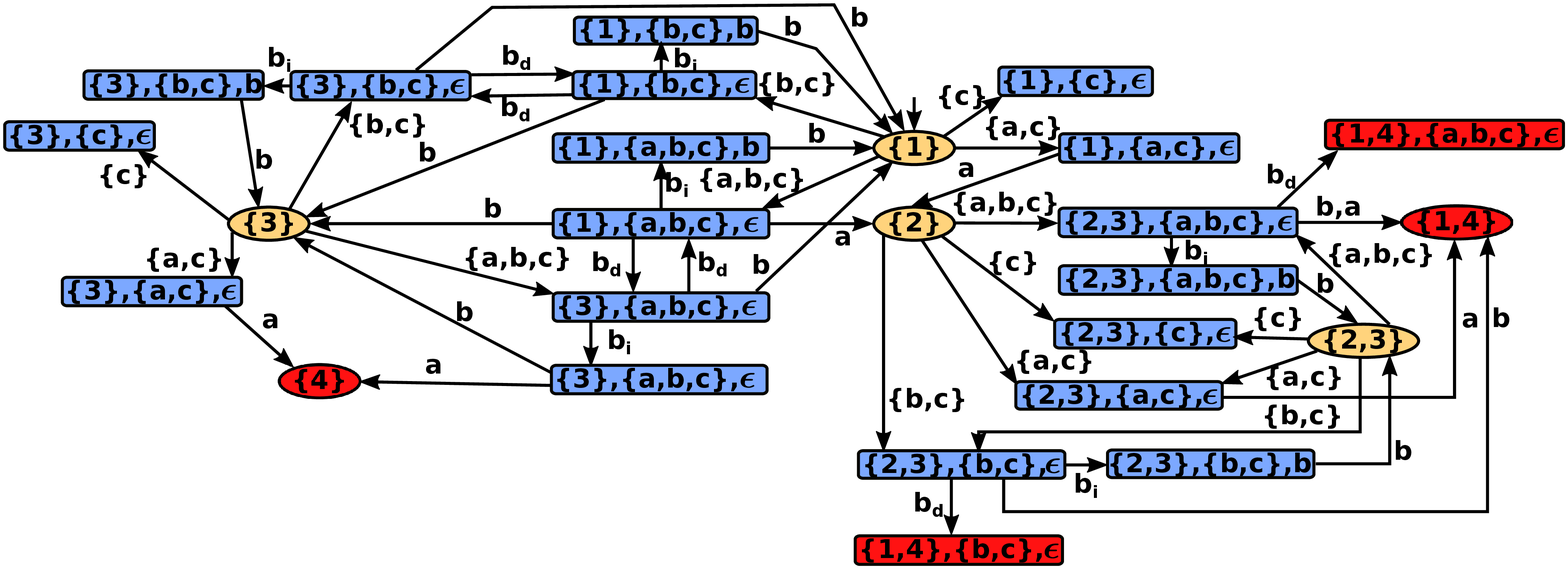}
      \caption{Full arena $\arenam$}
      \label{fig:arena}
\end{figure*}

\end{exmp}

For convenience, we extend the definition of $h_2$ based on a given control decision.
Namely, we define a transition function $H_2$ that always start and end in $Q_2$ states.
This notation simplifies walks in $\mathcal{A}$.

\begin{defn}\label{def:ind_q2} We define the function $H_2:Q_2\times \Sigma_{o,e} \times \Gamma \rightarrow Q_2$ as:
\begin{equation} \label{eq:H2}
\begin{aligned}
H_2(q,e,\gamma) \coloneqq \left\{
\begin{array}{ll}
	h_1(h_2(q,e),\gamma), &\text{if } e\in\Sigma_o\\
    h_2(q,e) &\text{if } e\in\Sigma^d_a\\
    h_1(h_2(h_2(q,e),\mathcal{M}(e)),\gamma), &\text{if } e\in\Sigma^i_a\\
	\text{undefined,}&\text{otherwise}
\end{array}
\right.
\end{aligned}
\end{equation}
The function $H_2$ can be recursively extended for strings $s\in\Sigma_{o,e}^*$ given a sequence of control decisions $\gamma_1\dots\gamma_{|s|}$, i.e., $H_2(q,s,\gamma_1\dots\gamma_{|s|}) = H_2(H_2(q,s^{|s|-1},\gamma_1\dots\gamma_{|s|-1}),e^{|s|}_s,\gamma_{|s|})$.
\end{defn} 


\subsection{Properties}

For a fixed supervisor $R$ and attacker $A$, we obtain the language $\lang(G_a||R_a||A)$ which contains the possible string executions in the attacked system, e.g., strings of events in $\Sigma_m = \Sigma\cup\Sigma_a^e$.
Given a string $s\in P_{\Sigma_m\Sigma_{o,e}}(\lang(G_a||R_a||A))$, we can find the state estimate of $G_a$ after execution of $s$, i.e., the state estimate of $G_a$ under supervision of $R_a$ and attack strategy $A$.
Formally, this state estimate is 
\begin{align}\label{eq:RE}
RE(s) =&
\{x\in X_{G_a} \mid x=\delta_{G_a}(x_{0,G_a},t) \text{ for}\nonumber\\
&t\in P^{-1}_{\Sigma_m\Sigma_{o,e}}(s)\cap\lang(G_a||R_a||A)\} 
\end{align}

In the construction of $\arenam$, we allow the attacker to insert events that are not allowed by the current control decision (see Eq.~(\ref{eq:h2_ins})).
Therefore, given a supervisor $R$, we need to define its control decisions for all $s\in \Sigma_o^*$, differing from the usual definition only for $s\in P_{\Sigma\Sigma_o}(\lang(G))$.
For this reason, we extend the function $\delta_R$ to be a complete function in $\Sigma_o$.
\begin{align}\label{eq:completedeltaR}
\Delta_R(x,e) &=\left\lbrace\begin{array}{ll}
\delta_R(x,e)&\text{if } e\in \Gamma_R(x)\\
x &\text{otherwise} \end{array} \right.  
\end{align}
for $x\in R$ and $e\in \Sigma_o$.
Intuitively, $\Delta_R$ extends $\delta_R$ by simply ignoring the events that are not defined in $\delta_R$.
The function $\Delta_R$ is extended to $s\in \Sigma^*_o$ as $\delta_R$ is extended.
Lastly, we define the control decision of $R$ for any $s\in \Sigma_o^*$ as:
\begin{equation}\label{eq:controlR}
\control_R(s) = \Gamma_R(\Delta_R(x_{0,R},s))
\end{equation}

Based on $H_2$ and $\control_R$, we show that the arena $\arenam$ computes the same state estimates based on the supervisor $R$ and attacker $A$ as the ones computed based on $G_a||R_a||A$.
This result is shown in Proposition~\ref{prop:runs_in_A} and its proof is in Appendix~\ref{app:proofs}.

\begin{prp}\label{prop:runs_in_A}
Given a system $G$, a supervisor $R$, an attack function $A$ and arena $\arenam$, then for any $s\in P_{\Sigma_m\Sigma_{o,e}}(\lang(G_a||R_a||A))$, we have that 
\begin{align}
H_2(x_0,s, \gamma_1\dots\gamma_{|s|})! \label{eq:runex}\\
I_1(H_2(x_0,s,\gamma_1\dots\gamma_{|s|})) &= RE(s)\label{eq:run1}\\
I_2(H_2(x_0,s,\gamma_1\dots\gamma_{|s|})) & = \delta_A(x_{0,A},s) \label{eq:run2}
\end{align}
where $x_0 = h_1(q_0,\control_R(\epsilon))$ and $\gamma_i = \control_R(P^S(s^i))$.
\end{prp}

Recall that in the left box of Fig.~\ref{fig:relationsys-meta} the supervisor is unknown.
Equation~(\ref{eq:runex}) tells us that the arena captures all possible interactions between any supervisor $R$ and attack function $A$ with the plant $G$.
It captures all possible interactions since Proposition~\ref{prop:runs_in_A} is true regardless of the supervisor $R$ and of the attack function $A$.
This is one of the main benefits of constructing the arena \arena . 
It defines a space where all supervisors and attacker actions based on $A$ for the plant $G$ exist.

Moreover, Eqs.~(\ref{eq:run1}-\ref{eq:run2}) says that the arena correctly captures the interaction between the attacker, supervisor and plant.
Equation~(\ref{eq:run1}) computes $G$'s state estimate of  based on the modified string $s\in P_{\Sigma_m\Sigma_{o,e}}(\lang(G_a||R_a||A))$ and the control decisions taken by $R$ along the observed string.
These estimates capture an agent that has full knowledge of the modification on the string $s$ and the decisions taken by $R$.
On the other hand, Eq.~(\ref{eq:run2}) establishes the correct state of the attacker $A$ in the construction of $\arenam$. 

The arena $\arenam$ has, in worst-case, $|X_A|2^{|X_G|}$ $Q_1$-states and $|X_A|(|\Sigma_a|+1)2^{|X_G|+|\Sigma_{c}|}$ $Q_2$-states given that $|\Gamma|\leq 2^{|\Sigma_c|}$.
Consequently, the worst-case running time of the construction of the arena $\arenam$ is $O(|X_A||\Sigma_o|^2 2^{|X_G|+|\Sigma_{c}|})$ since $\Sigma_a\subseteq \Sigma_o$.
We can construct $\arenam$ starting from its initial state and performing a breadth-first search based on equations $h_1$ and $h_2$.

\subsection{Solution of the Meta-Control problem}\label{subsection:meta}

Our approach to solve Problem~\ref{prob:synthesis} is to consider the above-constructed arena $\mathcal{A}$ as the \emph{uncontrolled system} in a \emph{meta-control problem}, which is posed as a supervisory control problem for a partially-observed discrete event system, as originally considered in \cite{Lin:1988}.
For that reason,  we will refer to $\mathcal{A}$ as the \textit{meta-system}. 
As will become clear in the following discussion, this supervisory control approach naturally captures our synthesis objectives, and moreover supervisory control theory provides a complete characterization of the solution.
Such a methodology was previously used in \cite{Yin:2016a,wu:2016b} for instance; however, in these works the meta-control problem is a control problem under full observation. 
The same situation does not apply in our case, where events in $\Sigma_a^e$ are unobservable (from the supervisor's perspective). 

To formally pose the meta-control problem, we need a specification for the meta-system. 
In fact, the specification emerges from the corresponding specification in Problem \ref{prob:synthesis}, which states that the controlled system should never reach any state in $X_{crit}$. 
The same specification is to be enforced in \arena, where the state estimate of $G$ represents the reachable states of $G$. 
Thus, the specification for the meta-control problem is that the meta-controlled system should never reach any state $q\in Q_1\cup Q_2$ such that $I_1(q)\cap X_{crit}\neq \emptyset$. 

The next step in the meta-control problem formulation is to specify the controllable and observable events in the meta-system \arena. 
We already mentioned that all $e\in\Sigma_a^e$ are unobservable events.
In fact, they are the only unobservable events in \arena\ since they are moves of the attacker that the supervisor does not directly observe. 
In regard to the controllable events, the supervisor makes decisions in order to react to the decisions made by the environment.
Therefore, the events in $A_1\setminus\{\Sigma_{uc}\}$ are controllable, while those in $A_2\cup\{\Sigma_{uc}\}$ are uncontrollable. 
Note that, we explicitly exclude the control decision composed only of uncontrollable events as a meta-controllable event; the supervisor should always be able to at least enable the uncontrollable events, otherwise it would not be admissible. 
In this way, the supervisor can always issue at least one control decision, i.e., enable all uncontrollable plant events. 
We are now able to formulate the meta-control problem.

\begin{defn}\label{def:meta-control} Given $\arenam$ constructed with respect to $G$ and $A$, with events $E = A_1\cup A_2$, $E_c= A_1\setminus\{\Sigma_{uc}\}$ as the set of controllable events and $E_{uo} = \Sigma_a^e$ as the set of unobservable events. 
Let \arena$^{trim} = trim($\arena$,M)$ be the specification automaton, where $M = \{q\in Q^{\mathcal{A}}_1\cup Q^{\mathcal{A}}_2| I_1(q)\cap X_{crit}\neq \emptyset\}$\footnote{We use superscripts to differentiate the different arena structures, e.g., $\arenam$, $\arenam^{trim}$, etc.}.
Calculate the supremal controllable and normal sublanguage of the language of \arena$^{trim}$ with respect to the language of \arena, and let this supremal sublanguage be generated by the solution-arena denoted by $\arenam^{\sup}$.
\end{defn}

Note that all controllable events in the meta-control problem are also observable, i.e., $E_c\subseteq E_o$. 
Therefore, the controllability and observability conditions are equivalent to the controllability and normality conditions.
Hence, in this case, the supremal controllable and observable sublanguage exists and is equal to the supremal controllable and normal sublanguage; see, e.g., \S 3.7.5 in \cite{Lafortune:2008}. 
\emph{As consequence a supremal and unique solution of the meta-control problem exists.}
This solution is the language generated by the solution-arena $\arenam^{\sup}$.

The state structure of $\arenam^{\sup}$ will depend on the algorithm used to compute the supremal controllable and normal sublanguage of \arena$^{trim}$.
One example of the structure of $\arenam^{\sup}$ is provided.  

\begin{exmp} \label{ex:Asup}
We return to our running example.
Based on $\arenam$, we obtain $\arenam^{\sup}$ using an integrated (for controllability and normality) iterative algorithm to compute the supremal controllable and normal sublanguage that is based on preprocessing the input automata to satisfy simultaneously a strict sub-automaton \cite{Lafortune:2008} condition and a State Partition Automaton \cite{Jirskova:2012} condition.
As part of the algorithm, one needs to refine $\arenam$ so that its observer is a state partition automaton (using algorithm in \cite{Jirskova:2012}), i.e., to compute $\arenam||Obs(\arenam)$, where $Obs$ is the observer operation with respect to $E_{uo}$ \cite{Lafortune:2008}.
The resulting $\arenam^{\sup}$ is depicted in Fig.~\ref{fig:Asup}.
Each state in $\arenam^{\sup}$ is a tuple, where the first component is a state in $\arenam$ and the second component is a state in $Obs(\arenam)$, where $obs_1 = \{(\{1\},\{b,c\},\epsilon),(\{1\},\{b,c\},b),(\{3\},\{b,c\},\epsilon),\allowbreak(\{3\},\{b,c\},b)\}$ and $obs_2 = \{(\{1\},\{c\},\epsilon),(\{3\},\{c\},\epsilon)\}$.
\begin{figure}[thpb]
       \centering
      \includegraphics[width=1\columnwidth]{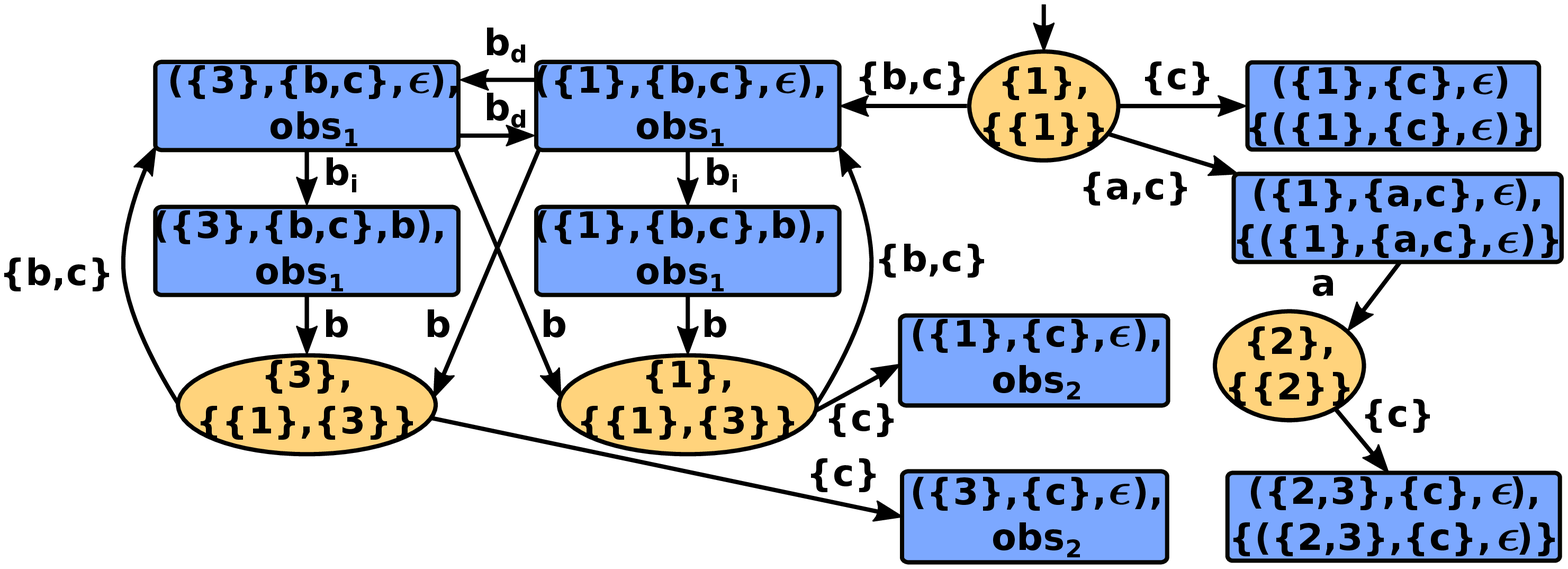}
      \caption{$\arenam^{\sup}$}
      \label{fig:Asup}
\end{figure}
\end{exmp}

Regardless of the algorithm to obtain $\arenam^{\sup}$, it has a structure with $Q_1$-like states and $Q_2$-like states, since it accepts a sublanguage of \arena$^{trim}$.
Namely, it has states where only control decisions are allowed ($Q_1$ states) and states where only transitions with events in $\Sigma_{o,e}$ are defined ($Q_2$ states).
Thus, we can use the functions previously defined for $\arenam$ in $\arenam^{\sup}$.

\begin{remk}
The worst-case running time to obtain the supremal controllable and normal sublanguage is exponential in product of the number of states of the system and the specification \cite{Brandt:1990}.
Therefore, the worst-case running time to obtain the $\arenam^{\sup}$ is $O(2^{(|Q_1|+|Q_2|)^2})$.
\end{remk}

The way \arena\ is constructed is such that it embeds the set of all supervisors for the original plant $G$.
Therefore,  the uniqueness of the language generated by \arena$^{\sup}$ and the fact that it is the supremal solution of the meta-control problem means that the structure \arena$^{\sup}$ embeds a family of supervisors $S$, where the controlled behavior generated by each member of that family does not reach any state in $X_{crit}$. 
Moreover, since \arena\ is constructed taking into account the attack function $A$,  this family of supervisors is robust with respect to $A$.
This leads us to the following result.
Its proof is in Appendix~\ref{app:proofs}.

\begin{thm}\label{prop:S_PinAsup}
A supervisor $R$ is a robust supervisor with respect to $A$ if and only if $(\forall s \in P_{\Sigma_m\Sigma_{o,e}}(\lang(G_a||R_a||A)))[H_2^{\arenam^{\sup}}(x_0,s,\gamma_1\dots\gamma_{|s|})!]$,
where $x_0 = h_1(q^{\arenam^{\sup}}_0,\control_R(\epsilon))$ and $\gamma_i = \control_R(P^S(s^i))$.  
\end{thm}

\begin{corollary}\label{theo:existence}
\arena$^{\sup} = \emptyset$ if and only if there does not exist any robust supervisor $R$ with respect to attacker $A$.
\end{corollary}
Theorem~\ref{theo:existence} states that a supervisor is robust if and only if it is embedded in $\arenam^{\sup}$.
Next, Corollary \ref{theo:existence} gives a necessary and sufficient condition for the existence of a solution for Problem \ref{prob:synthesis}.
Given that there exists a robust supervisor, we provide an algorithm\footnote{There are different manners for a designer to extract a robust supervisor.} to extract a supervisor that solves Problem~\ref{prob:synthesis}.
First, we define function $H_1$ as we defined $H_2$.

\begin{defn}\label{def:ind_q1} 
Let the function $H_1:Q_1\times \Sigma_{o,e} \times \Gamma \rightarrow Q_1$ be defined as:
\begin{equation} \label{eq:H1}
\begin{aligned}
H_1(q,e,\gamma) \coloneqq \left\{
\begin{array}{ll}
	h_2(h_1(q,\gamma),e), &\text{if } e\in\Sigma_o\\
    q &\text{if } e\in\Sigma^d_a\\
    h_2(h_2(h_1(q,\gamma),e),\mathcal{M}(e)), &\text{if } e\in\Sigma^i_a\\
	\text{undefined,}&\text{otherwise}
\end{array}
\right.
\end{aligned}
\end{equation}
\end{defn} 

\begin{algorithm}
\caption{Robust Supervisor Extraction}\label{algo:constructionRr}
\begin{center}
\begin{algorithmic}[1]
\Require $\arenam^{\sup}$
\Ensure $R_r = (X_{R_r},\Sigma,\delta_{R_r},x_{0,R_r})$
\State $x_{0,R_r} = q^{\arenam^{\sup}}_0$ 
\State $X_{R_r} \leftarrow \{x_{0,R_r}\}$, $\delta_{R_r} \leftarrow \emptyset$
\State Expand($x_{0,R_r}$)
\Procedure{Expand}{$x$}
\State select $\gamma \in \Gamma_{\arenam^{\sup}}(x)$ s.t. $\forall \gamma'\in \Gamma_{\arenam^{\sup}}(x): \gamma\not\subset\gamma'$
\ForAll{$e\in\Sigma\cap\gamma$}
\If{$e\in\Sigma_o$}
\State $y = H_1^{\arenam^{\sup}}(x,e,\gamma)$, $\delta_{R_r} \leftarrow \delta_{R_r} \cup (x,e,y)$
\State $X_{R_r} \leftarrow X_{R_r}\cup \{y\}$
\If{$y \notin X_{R_r}$}
\State $Expand(y)$
\EndIf
\Else
\State $\delta_{R_r} \leftarrow \delta_{R_r} \cup (x,e,x)$
\EndIf
\EndFor
\EndProcedure
\end{algorithmic}
\end{center}
\end{algorithm}

Algorithm~\ref{algo:constructionRr} starts at the initial state of \arena$^{\sup}$ and performs a Depth First Search by selecting the largest control decisions at each state that it visits.
By largest, we mean that it selects a control decision that is not a subset of any other control decision defined at state $x$, as described by line $5$.
Note that, it is possible to have more than two decisions that satisfy this condition.
In this case, the algorithm selects one of the possible decisions in a nondeterministic manner.
The algorithm terminates since $\arenam^{\sup}$ is finite.
Moreover, the algorithm only traverses player $1$ states, where the control decisions are defined.
\begin{corollary}\label{theo:synthesis}
A supervisor $R_r$ constructed by Algorithm \ref{algo:constructionRr} is a solution for Problem~\ref{prob:synthesis}.
\end{corollary}

\begin{remk}
The worst-case running time of Algorithm~\ref{algo:constructionRr} is linear in the number of state of $\arenam^{\sup}$.
For this reason, the running time of the entire synthesis procedure is exponential in the number of states of $\arenam$.
Since the number of states of $\arenam$ is exponential in the number of states of $G$, the overall worst-case running time is double exponential in the number of states of $G$, which is one exponential order smaller than in \cite{Su:2018} and one exponential order higher than in \cite{Meira-Goes:2019c}, two references that were reviewed in Section I.
\end{remk}

\begin{exmp} \label{ex:supalgo1}
To conclude this section, we provide two supervisors extracted via Algorithm~\ref{algo:constructionRr}.
These two supervisors are depicted in Fig.~\ref{fig:supervisors}.

\begin{figure}[thpb]
	\centering
	\begin{subfigure}[t]{.45\columnwidth}
		\centering
		\includegraphics[width=.5\columnwidth]{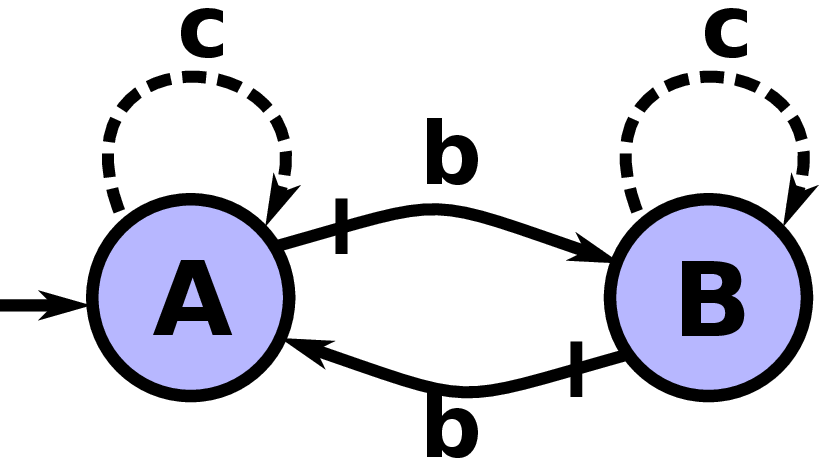}
		\caption{Robust supervisor $R_1$}
		\label{fig:sup1}
	\end{subfigure}%
	\quad
	\begin{subfigure}[t]{.45\columnwidth}
		\centering
		\includegraphics[width=0.5\columnwidth]{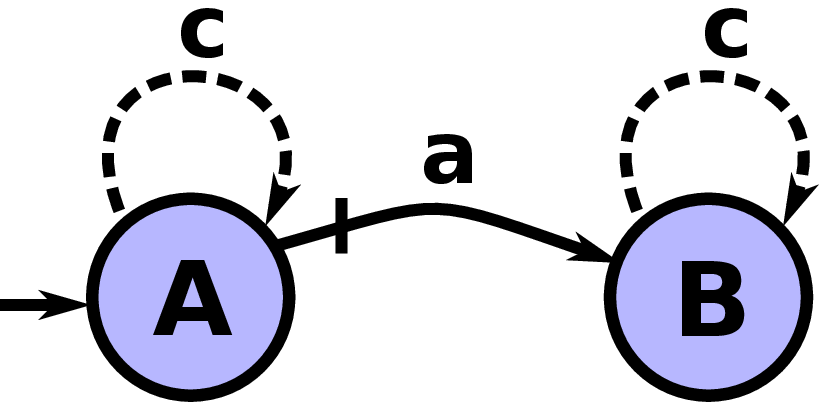}
		\caption{Robust supervisor $R_2$}
		\label{fig:sup2}
	\end{subfigure}
	\caption{Robust supervisors with respect to $A^1$}
	\label{fig:supervisors}
\end{figure}
\end{exmp}

\section{Selecting supervisors in the robust arena}\label{sec:supselect}

Algorithm~\ref{algo:constructionRr} provides one way of extracting robust supervisors from $\arenam^{\sup}$.
As we explained before, it selects maximal control decisions in the $Q_1$ states that the algorithm visits.
Example~\ref{ex:supalgo1} shows that this extraction does not provide specific information about the language generated by supervised system once a supervisor is selected, other than the fact that we are choosing a locally maximal control decision.
While supervisor $R_1$ in Fig.~\ref{fig:supervisors}(\subref{fig:sup1}) generates a live language, supervisor $R_2$ in Fig.~\ref{fig:supervisors}(\subref{fig:sup2}) is blocking.
Nonetheless, the space defined in $\arenam^{\sup}$ provides maximum flexibility in extracting different supervisors since all robust supervisors are embedded in $\arenam^{\sup}$.
The methods in \cite{Wakaiki:2017,Su:2018,Meira-Goes:2019c} do not provide the flexibility of $\arenam^{\sup}$ since they exploit algorithms of Supervisory Control Theory where only \emph{one} supervisor can be obtained at a time.
In fact, when explicit comparisons can be made, the supervisors obtained by their methods are embedded in the corresponding $\arenam^{\sup}$.

Another benefit of the construction of $\arenam^{\sup}$ is the ability to exploit results in the area of turn-based two-player graph-games.
Results from these areas can be leveraged to study different manners of extracting robust supervisors, e.g., to study quantitative versions of the robust supervisor problem under some cost model \cite{Cassez:2012,Ji:2018,Meira-Goes:2019d}.

We provide an example of a supervisor extraction algorithm based on a quantitative measure.
First, we define a measure over the supervised system $R/G$, i.e., over the states of the automaton $G||R$.
Let the set $X_{dead} = \{x \in X_{G||R}\mid \Gamma_{G||R}(\delta_{G||R}(x,s))=\emptyset \text{ for } s\in \Sigma_{uo}^*\}$ be the set of states in $G||R$ that can reach a deadlock state via an unobservable string.
We define $r:X_{G||R}\rightarrow [0,+\infty)\cup \{-\infty\}$ to be a reward function for any $(x,y)\in X_{G||R}$ and $c\in [0,\infty)$ as:
\begin{equation}\label{eq:reward}
r((x,y)) = \left\lbrace
\begin{array}{ll}
-\infty & \text{if } x\in X_{crit}\\
0 & \text{if } (x,y) \in X_{dead}\\
c & \text{otherwise}
\end{array}\right.
\end{equation}

The reward function $r$ punishes states from where the system $G||R$ might deadlock.
Based on the reward function $r$, we define the following total reward for the supervised system $G||R$.
\begin{equation}\label{eq:rewardsystem}
Reward(R,G) = \sum_{x\in X_{G||R}} r(x)
\end{equation}
We can generalize Algorithm~\ref{algo:constructionRr} to incorporate this quantitative measure such that it extracts a supervisor from $\arenam^{\sup}$ that maximizes the measure $Reward(R,G)$.
In our running example, this new method extracts supervisor $R_1$.
Further, we can assume that the attacker tries to minimize $Reward(R,G)$ in this extraction method.
In this scenario, we would pose a $\min\max$ problem in order to select a supervisor from $\arenam^{\sup}$.
We leave these extensions for future work.

\section{Robot Motion Planning Example} \label{sec:ex}
We developed a tool\footnote{Our software tool is available at: URL. URL will be included upon final acceptance of the paper.} to automatically construct $\arenam$, as in Definition~\ref{def:arena}, and to compute $\arenam^{\sup}$.
Moreover, Algorithm~\ref{algo:constructionRr} is also implemented in our tool.
Our evaluation was done on a Linux machine with 2.2GHz CPU and 16GB memory.

\begin{figure}[thpb]      
      \centering
	\begin{subfigure}[t]{0.45\columnwidth}
		\centering
       	\includegraphics[width=.8\columnwidth]{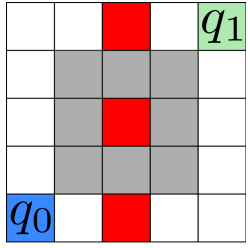}
      	\caption{Robot workspace: the robot starts in the blue cell denoted by $q_0$; the shaded area is considered hostile and the sensor readings in this area are compromised.}
      	\label{fig:part}
	\end{subfigure}%
	\quad
	\begin{subfigure}[t]{0.6\columnwidth}
		\centering
      	\includegraphics[width=1\columnwidth]{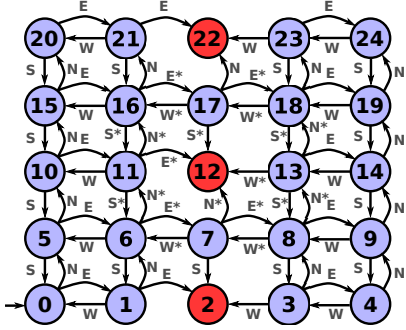}
      	\caption{Model of the robot in the workspace: $\Sigma = \Sigma_c = \Sigma_o = \{E,W,N,S,E*,W*,N*,S*\}$}
      	\label{fig:sim_rob_attack}
	\end{subfigure}
	\caption{Robot workspace and attack simulation}
	\label{fig:plane}
\end{figure}

We consider a robot moving in a possibly hostile environment.
The robot is assumed to have four different movement modes that are modeled as controllable and observable events.
The robot moves freely in the workspace shown in Fig.~\ref{fig:plane}(\subref{fig:part}).
Its initial state is the blue cell denoted as $q_0$ and the red cells are considered to be obstacles.
Moreover, the shaded region is assumed to be hostile and the sensor readings of the robot could be under attack.
This uncontrolled system is modeled by the automaton depicted in Fig.~\ref{fig:plane}(\subref{fig:sim_rob_attack}).
We want to design a robust supervisor that enforces the following properties: (1) the robot must avoid the obstacles; (2) the robot can always access states $q_0$ and $q_1$.

The set of compromised events is $\Sigma_a = \{E*,W*,N*,S*\}$ since we consider that the sensor readings in the shaded area might be under attack.
First, we construct the arena $\arenam$ considering the all-out attack strategy.
The number of states in $\arenam$ is $18649$ states.
The state space explosion is due to the number of control decisions: there are $256$ possible control decisions.

After constructing $\arenam$, we obtain $\arenam^{\sup}$ as described in Definition~\ref{def:meta-control}.
To compute the supremal controllable and normal sublanguage, we used the algorithm described in Example~\ref{ex:Asup}.
The number of states in $\arenam^{\sup}$ is $65358$ states. 
Note that $\arenam^{\sup}$ has more states than $\arenam$. 
The larger state space in $\arenam^{\sup}$ is due to necessary preprocessing done by the iterative algorithm for the computation of the supremal controllable and normal sublaguage.

\begin{figure}[thpb]
      \centering
      \includegraphics[width=0.7\columnwidth]{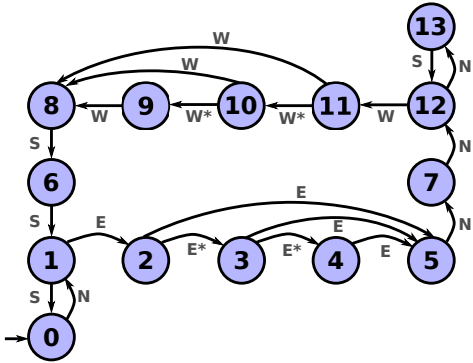}
      \caption{Robust supervisor for robot in a hostile environment}
      \label{fig:suprobot}
\end{figure}

Finally, any supervisor selected from $\arenam^{\sup}$ is robust against the all-out attack strategy, i.e., it satisfies property (1).
Therefore, we must select a supervisor that satisfies property (2).
For this reason, we modify Algorithm~\ref{algo:constructionRr} to extract a supervisor from $\arenam^{\sup}$ such that property (2) is satisfied.
This supervisor is depicted in Fig.~\ref{fig:suprobot}.

\section{Conclusion}\label{sec:conclusion}
We have considered a class of problems in cyber-security where sensor readings in a feedback control system may be manipulated by a malicious attacker.
By formulating the problem at the supervisory control layer of a cyber-physical system, we were able to leverage techniques from games on automata under partial information and from supervisory control of partially-observed discrete event systems to develop a solution methodology to prevent damage to the system when some sensor readings may be edited by the attacker.
Our problem formulation is parameterized by an attacker strategy over a set of compromised events.
In this manner, synthesis of robust supervisors against sensor deception attack strategies is considered, e.g., bounded attack strategies, replacement attack strategies, etc.  
Moreover, if there is no prior information about the attacker strategy, then we consider the general all-out attack strategy.
A supervisor robust against the all-out attack strategy is robust against any other sensor deception attack strategy.

The space defined in $\arenam^{\sup}$ provides maximum flexibility in extracting different supervisors since all robust supervisors are embedded in $\arenam^{\sup}$.
As discussed in Section~\ref{sec:supselect}, it would be interesting to investigate methods to extract supervisors from \arena$^{\sup}$ in order to satisfy additional constraints, such as optimality with respect to some quantitative criterion \cite{Ji:2018,Cassez:2012}.
Finally, identifying ways to reduce the state space of the arena by exploiting a suitable notion of state equivalence is another important research direction.

\appendices

\section{Attack function encoding}\label{app:attack}
In Section~\ref{sec:robust}, we assume that the automaton $A$ that encodes an attack function $f_A$ has a transition function complete with respect to $\Sigma_o\setminus\Sigma_a$ and for any $(e\in\Sigma_a,\ q\in X_A)$ then $(\delta_A(q,e)!\vee\delta_A(q,e_d)!)$.
To encode any attack function $f_A$ as an automaton, this assumption does not need to be satisfied.
However, this means that the attacker might ``block" the controlled system if it receives an unexpected event executed by the plant, i.e., $f_A(s,e)$ is undefined. 
 
Based on the above assumption, we relax the completeness assumption of $\delta_A$ in order to encode any attack function $f_A$ as an automaton.
Namely, the partial transition function $\delta_A$ of $A$ must satisfy one of the following conditions for any $q\in X_A$:
\begin{enumerate}
\item[(1)] $(\forall e\in\Sigma_o\setminus\Sigma_a)[\delta_A(q,e)!]$ and $(\forall e\in\Sigma_a)[\delta_A(q,e)!\vee\delta_A(q,e_d)!]$; or
\item[(2)] $(\forall e\in\Sigma_o\cup\Sigma_a^d)[\delta_A(q,e)$ is not defined$]$ and $(\exists e\in\Sigma_a^i)[\delta_A(q,e)!]$;
\end{enumerate}

Condition (2) allows attack strategies where event insertion is faster than the plant executes events.
Note that condition (2) violates controllability since it temporarily blocks the plant from executing events.
The practicality of this assumption will depend on the plant's response time, i.e., this attack condition is application dependent.
Condition (1) remains unchanged, i.e., the attacker does not block the plant in states that satisfy this condition.
\section{Proofs}\label{app:proofs}
Proposition~\ref{prop:runs_in_A}
\begin{proof}
The result is proved by induction on the length of the string $s \in P_{\Sigma_m\Sigma_{o,e}}(\lang(G_a||R_a||A))$.

Before we start the induction proof, we state two important results.
First, we define $\control_{R_a}$ in the same manner as $\control_{R}$, but the control decisions of $R_a$ are defined over $\Sigma_m$.
It can be shown by induction that the following equality holds for any $s\in \lang(G_a||R_a||A)$:
\begin{equation}\label{eq:R_R_a}
\control_R(P^S(s)) = \control_{R_a}(s)\cap \Sigma
\end{equation}
Intuitively, Eq.~(\ref{eq:R_R_a}) follows since $R_a$ is a copy or $R$ with insertion and deletions events added based on $P^S$. 

Second, the function $RE$ can also be computed recursively as follows for $s\in P_{o,e}(\lang(G_a||R_a||A))$ and $e\in \Sigma_{o,e}$:
\begin{align}\label{eq:RErec}
RE(se) =& \{x\in X_{G_a}\mid x = \delta_{G_a}(\delta_{G_a}(RE(s),e),t)\text{ for} \nonumber\\
&t\in (\Sigma_{uo}\cap\control_{R_a}(se))^*\}
\end{align}
$\Sigma_{uo}$ defines the unobservable events of $G_a$.
Then, only supervisor $R_a$ disables events in $\Sigma_{uo}$ to be executed in $G_a$ since $A$ is defined over $\Sigma_{o,e}$.
For this reason, Eq.~\ref{eq:RErec} is equivalent to Eq.~\ref{eq:RE}.

Induction basis: $s = \epsilon$.

We have that $H_2(x_0,\epsilon,\epsilon) = h_1(q_0,\control_R(\epsilon))$ is well defined since $h_1$ is complete with respect to $\Gamma$.
It also follows that $\delta_A(x_{0,A},\epsilon) = I_2(x_0)$ since $\epsilon \in \lang(G_a||R_a||A)$ and $I_2(x_0) = x_{0,A}$.
We have that $I_1(H_2(x_0,\epsilon,\epsilon)) = UR_{\control_{R}(\epsilon)}(x_{0,G})$.
\begin{align}
RE(\epsilon) \smash{\overset{\textrm{Eq.(\ref{eq:RE})}}{=}}& \{x\in X_{G_a}\mid x = \delta_{G_a}(x_{0,G_a},t)\text{ for} \nonumber\\
&t\in P^{-1}_{\Sigma_m\Sigma_{o,e}}(\epsilon)\cap\lang(G_a||R_a||A)\}\\
\overset{\substack{\mathrm{Def.\ref{def:G_a}}\\ \mathrm{P_{\Sigma_m\Sigma_{o,e}}}}}{=}& \{x\in X_{G} \mid x = \delta_{G}(x_{0,G},t)\text{ for} \nonumber\\
&t\in \Sigma_{uo}^*\cap\lang(G_a||R_a||A)\}\\
\overset{\substack{\mathrm{Def.\ref{def:R_a}}}}{=}& \{x\in X_{G} \mid x = \delta_{G}(x_{0,G},t)\text{ for}\nonumber \\
&t\in (\Sigma_{uo}\cap\control_{R_a}(\epsilon))^*\}\\
\overset{\substack{\mathrm{Eq.(\ref{eq:R_R_a})}}}{=}& \{x\in X_{G} \mid x = \delta_{G}(x_{0,G},t)\text{ for}\nonumber \\
&t\in (\Sigma_{uo}\cap\control_{R}(\epsilon))^*\}\\
\overset{\substack{\mathrm{Eq.(\ref{eq:UR})}}}{=}& UR_{\control_R(\epsilon)}(x_{0,G})
\end{align}

Induction hypothesis: $H_2(x_0,s,\gamma_1\dots\gamma_{|s|})!$, $I_1(H_2(x_0,s,\gamma_1\dots\gamma_{|s|})) = RE(s)$ and $I_2(H_2(x_0,s,\gamma_1\dots\gamma_{|s|})) = \delta_A(x_{0,A},s)$ for all $s\in \lang(G_a||R_a||A)$ and $|s| = n$.

Induction step: 
Let $e\in \Sigma_{o,e}$, $s\in \lang(G_a||R_a||A)$, $|s| = n$ and $se\in \lang(G_a||R_a||A)$.
The induction hypothesis gives us $H_2(x_0,s,\gamma_1\dots\gamma_{|s|})!$, $I_1(H_2(x_0,s,\gamma_1\dots\gamma_{|s|})) = RE(s)$, and $I_2(H_2(x_0,s,\gamma_1\dots\gamma_{|s|})) = \delta_A(x_{0,A},s)$.
Let $q = H_2(x_0,s,\gamma_1\dots\gamma_{|s|})$.

Since $se\in \lang(G_a||R_a||A)$ then it follows that $H_2(q,e,\gamma_{|se|})!$.
Moreover, it follows that $I_2(H_2(x_0,se,\gamma_1\dots\gamma_{|se|})) = \delta_A(x_{0,A},se)$ by construction of $\arenam$.

For equality of Eq.~(\ref{eq:run1}), we divide the event $e$ into three cases.

First, $e\in\Sigma_a^d$.
Then, $\gamma_{|se|} =  \gamma_{|s|}$
Based on the construction of $\arenam$, we have that $I_1(H_2(q,e,\gamma_{|se|})) = UR_{\gamma_{|s|}}(NX_{P^G(e)}(I_1(q)))$.
\begin{align}
RE(se) \overset{\substack{\mathrm{Eq.(\ref{eq:RErec})}}}{=}& \{x\in X_{G_a} \mid x = \delta_{G_a}(\delta_{G_a}(RE(s),e),t)\text{ for} \nonumber\\
&t\in (\Sigma_{uo}\cap\control_{R_a}(se))^*\}\\
\overset{\substack{\mathrm{Def.\ref{def:G_a}}\\\mathrm{Eq.(\ref{eq:R_R_a})}}}{=}& \{x\in X_{G} \mid x = \delta_{G}(\delta_{G}(RE(s),P^G(e)),t)\text{ for} \nonumber\\
&t\in (\Sigma_{uo}\cap\control_{R}(P^S(se)))^*\}\\
\overset{\substack{\mathrm{Eq.(\ref{eq:nx})}\\\mathrm{\gamma_{|s|}= \gamma_{|se|}}}}{=}& \{x\in X_{G} \mid x = \delta_{G}(NX_{P^G(e)}(RE(s)),t)\text{ for} \nonumber\\
&t\in (\Sigma_{uo}\cap\gamma_{|s|})^*\}\\
\overset{\substack{\mathrm{Eq.(\ref{eq:UR})}}}{=}& UR_{\gamma_{|s|}}(NX_{P^G(e)}(RE(s))) \\
= U&R_{\gamma_{|s|}}(NX_{P^G(e)}(I_1(q)))
\end{align}


Let, $e\in\Sigma_a^i$.
Based on the construction of $\arenam$, we have that $I_1(H_2(q,e,\gamma_{|se|})) = UR_{\gamma_{|se|}}(I_1(q))$.
\begin{align}
RE(se) \overset{\substack{\mathrm{Eq.(\ref{eq:RErec})}}}{=}& \{x\in X_{G_a} \mid x = \delta_{G_a}(\delta_{G_a}(RE(s),e),t)\text{ for} \nonumber\\
&t\in (\Sigma_{uo}\cap\control_{R_a}(se))^*\}\\
\overset{\substack{\mathrm{Def.\ref{def:G_a}}\\\mathrm{Eq.(\ref{eq:R_R_a})}}}{=}& \{x\in X_{G} \mid x = \delta_{G}(\delta_{G}(RE(s),P^G(e)),t)\text{ for} \nonumber\\
&t\in (\Sigma_{uo}\cap\control_{R}(P^S(se)))^*\}\\
\overset{\substack{P^G(e) = \epsilon}}{=}& \{x\in X_{G} \mid x = \delta_{G}(RE(s),t)\text{ for} \nonumber\\
&t\in (\Sigma_{uo}\cap\gamma_{|se|})^*\}\\
\overset{\substack{\mathrm{Eq.(\ref{eq:UR})}}}{=}& UR_{\gamma_{|se|}}(RE(s)) \\
= U&R_{\gamma_{|se|}}(I_1(q))
\end{align}

Lastly, $e\in\Sigma_o$.
Based on the construction of $\arenam$, we have that $I_1(H_2(q,e,\gamma_{|se|})) = UR_{\gamma_{|se|}}(NX_{e}(I_1(q)))$.
\begin{align}
RE(se) \overset{\substack{\mathrm{Eq.(\ref{eq:RErec})}}}{=}& \{x\in X_{G_a} \mid x = \delta_{G_a}(\delta_{G_a}(RE(s),e),t)\text{ for} \nonumber\\
&t\in (\Sigma_{uo}\cap\control_{R_a}(se))^*\}\\
\overset{\substack{\mathrm{Def.\ref{def:G_a}}\\\mathrm{Eq.(\ref{eq:R_R_a})}}}{=}& \{x\in X_{G} \mid x = \delta_{G}(\delta_{G}(RE(s),P^G(e)),t)\text{ for} \nonumber\\
&t\in (\Sigma_{uo}\cap\control_{R}(P^S(se)))^*\}\\
\overset{\substack{\mathrm{Eq.(\ref{eq:nx})}\\ P^G(e) = e}}{=}& \{x\in X_{G} \mid x = \delta_{G}(NX_{e}(RE(s)),t)\text{ for} \nonumber\\
&t\in (\Sigma_{uo}\cap\gamma_{|se|})^*\}\\
\overset{\substack{\mathrm{Eq.(\ref{eq:UR})}}}{=}& UR_{\gamma_{|se|}}(NX_{e}(RE(s))) \\
= U&R_{\gamma_{|se|}}(NX_{e}(I_1(q)))
\end{align}

This concludes our proof.
\end{proof}

Theorem~\ref{theo:existence}
\begin{proof}
We start with the only if part. 
Let $R$ be a robust supervisor.
Proposition~\ref{prop:runs_in_A} guarantees that $H_2^{\arenam}(x_0,s,\gamma_1\dots\gamma_{|s|})$ is defined for any $s \in \lang(G_a||R_a||A)$ and for any attack function representation $A$.
To analyze the meta-system $\arenam$, we have to define some notation for it.

We are analyzing $\arenam$ as a meta-system, namely as an automaton.
The function $h$ is a the combination of the functions $h_1$ and $h_2$.
Let $E = A_1\cup A_2$ be the event set of $\arenam$, $E_o = E\setminus\Sigma_e^a$ the observable event set, $E_{c} = A_1\setminus\{\Sigma_{uc}\}$ the controllable event set.
Moreover, the function $\eta:E^*\rightarrow \Sigma^*_o$ projects strings in $E^*$ to strings in $\Sigma^*_o$.
Intuitively, for any $s\in \lang(\arenam)$ the function $\eta(s)$ returns the string that is observed by the supervisor.

We construct the language $L\subset\lang(\arenam)$ recursively as:
\begin{enumerate}
\item $\epsilon \in L$
\item $s\in L\wedge h(q_0,s)\in Q^{\arenam}_2\Rightarrow se\in L$, $\forall e\in \Gamma_{\arenam}(h(q_0,s))$
\item $s\in L\wedge h(q_0,s)\in Q^{\arenam}_1 \wedge \big(e = \{\Sigma_{uc}\} \vee e = \control_R(\eta(s))\big) \Rightarrow se\in L$ 
\end{enumerate}
The language $L$ is by construction controllable w.r.t. $E_c$ and $\lang(\arenam)$; we show that $L$ is normal w.r.t. $E_o$ and $\lang(\arenam)$.
The result is shown by contradiction.
Assume that $L$ is not normal, then there exist shortest $s\in L$ and $t \in \lang(\arenam)\setminus L$ s.t. $P(s) = P(t)$.
In the construction of $L$, player 2 is not constrained, meaning that the shortest strings that belong to $\lang(\arenam)\setminus L$ end with an event in $A_1$ (control decisions).
For this reason, $e^{|s|}_s = e^{|t|}_t$ and $P_{EE_o}(s^{|s|-1}) = P_{EE_o}(t^{|t|-1})$. 
It implies that $\eta(s^{|s|-1}) = \eta(t^{|t|-1})$ and $\control_R(\eta(s^{|s|-1})) = \control_R(\eta(t^{|t|-1}))$.
By the definition of $L$, $t\in L$.
This contradicts our assumption.

It is also true that $L\subseteq \lang(\arenam^{trim})$, otherwise $R$ would not be a robust supervisor.
Intuitively, the actions made by player 2 are not constrained in the construction of $L$.
This guarantees that $L$ embeds all actions of attacker $A$.
The actions of player 1 are constrained based on $R$ and $\{\Sigma_{uc}\}$.
$R$ is robust and changing any of its control actions for any string by $\{\Sigma_{uc}\}$ will preserve robustness since $\{\Sigma_{uc}\}\subseteq \gamma$ for any $\gamma \in \Gamma$.

Definition \ref{def:meta-control} defines $\lang(\arenam^{\sup})$ to be the supremal controllable and normal sublanguage of $\lang(\arenam^{trim})$ w.r.t. $E_c,E_o$ and $\lang(\arenam)$.
Since $L\subseteq \lang(\arenam^{trim})$ and it is controllable and normal, then $L\subseteq \lang(\arenam^{\sup})$.
Therefore, $H_2^{\arenam^{\sup}}(x_0,s,\gamma_1\dots\gamma_{|s|})!$ holds for all $s\in P_{\Sigma_m\Sigma_{o,e}}(\lang(G_a||R_a||A))$.

For the if part, the result follows from the construction of $\arenam$, $\arenam^{trim}$ and the properties of $\arenam^{\sup}$. 
\end{proof}

\section*{Acknowledgment}
It is a pleasure to acknowledge many useful discussions with Lo\"ic  H\'elou\"et in the preparation of this paper.
The authors are also grateful to the reviewers for their insightful comments.


%

\ifCLASSOPTIONcaptionsoff
  \newpage
\fi



%

\bibliographystyle{IEEEtran}        
\bibliography{IEEEbib}           

\end{document}